\PassOptionsToPackage{colorlinks=true,linkcolor=blue,citecolor=blue}{hyperref}
\documentclass[prd,twocolumn,nofootinbib,superscriptaddress,letterpaper]{revtex4}

\newcommand\ForInternalReference[1]{}
\newcommand\SkipForEarlyCirculation[1]{}

\newcommand\SkipPP[1]{}
\usepackage{tabularx}
\usepackage{array}
\usepackage{graphicx}
\usepackage{dcolumn}
\usepackage{bm}
\usepackage{xspace}
\usepackage{url}
\usepackage{amsmath}
\usepackage{longtable}
\usepackage{listings}
\usepackage{multirow}
\usepackage{amssymb}
\usepackage{color}
\usepackage{tikz}
\usepackage{orcidlink}
\usepackage{cleveref}
\usepackage{xurl}
\usetikzlibrary{shapes.geometric, arrows, bayesnet}
\usepackage{hyperref}
\newcommand\optional[1]{}

\tikzstyle{startstop} = [circle, rounded corners, minimum width=1cm, minimum height=1cm,text centered, draw=black, fill=red!30]
\tikzstyle{io} = [trapezium, trapezium left angle=70, trapezium right angle=110, minimum width=2cm, minimum height=1cm, text centered, draw=black, fill=blue!30]
\tikzstyle{process} = [rectangle, minimum width=2cm, minimum height=1cm, text centered, draw=black, fill=orange!30]
\tikzstyle{decision} = [diamond, minimum width=2cm, minimum height=1cm, text centered, draw=black, fill=green!30]
\tikzstyle{arrow} = [thick,->,>=stealth]
\definecolor{amber}{rgb}{1.0, 0.75, 0.0}
\definecolor{orange}{rgb}{1.0, 0.5, 0.0}
\definecolor{amaranth}{rgb}{0.9, 0.17, 0.31}

\graphicspath{{./figures/}}

\def\ltsima{$\; \buildrel < \over \sim \;$}
\def\simlt{\lower.5ex\hbox{\ltsima}}
\def\gtsima{$\; \buildrel > \over \sim \;$}
\def\simgt{\lower.5ex\hbox{\gtsima}}

\newcommand\gwk{\textsc{GWKokab}\xspace}

\newcommand{\pvec}{\Lambda} 
\newcommand{\svec}{\lambda} 
\newcommand{\comp}{\boldsymbol{\rho}}
\newcommand{\svecz}{\boldsymbol{\lambda}} 
\newcommand{\pvecz}{\boldsymbol{\Lambda}}
\newcommand{\data}{\mathcal{D}}

\def\RIT{Center for Computational Relativity and Gravitation, Rochester Institute of Technology, Rochester, New York 14623, USA}

\begin{document}

\renewcommand{\arraystretch}{1.5}
\title{Population Properties of Binary Black Holes with Eccentricity}

\author{M. Zeeshan\orcidlink{0000-0002-6494-7303}}
\email{m.zeeshan5885@gmail.com}
\affiliation{\RIT}

\author{R. O'Shaughnessy\orcidlink{0000-0001-5832-8517}}
\affiliation{\RIT}

\author{N. Malagon\orcidlink{0000-0002-5825-7795}}
\affiliation{\RIT}

\begin{abstract}

The development of eccentric waveform models enables us to explore the growing catalog of gravitational-wave events with measurable eccentricity. This opens new opportunities to gain insight into the formation channels and evolutionary pathways of compact binary systems using eccentricity. However, most recent population analyses have been limited to quasi-circular binaries, primarily due to constraints in waveform modeling and sensitivity estimates. We are now entering an era where both of these limitations are being addressed, allowing for a more comprehensive investigation of eccentric binary populations. In this work, we
perform a very first population analysis that simultaneously fits the mass, spin, redshift, and eccentricity distribution. Specifically, we use source-parameter estimation on 153 binary black holes in GWTC-4 catalog provided by the Rapid Iterative FiTting (RIFT) framework using the \textsc{SEOBNRv5EHM} waveform model. We extend the default O4a population model to include orbital
eccentricity. We find that inferred population properties are broadly
consistent with conclusions obtained in previous analyses assuming quasi-circular binaries.
To assess sensitivity of our results to the most eccentric sources, we repeat our analysis excluding GW200129\_065458.
Consistent with our conclusions about each event and using \textsc{Nonoverlapping Mixture} eccentricity model, we bound
the branching ratio for eccentric events to be below $0.051890$ and $0.022011$ at $90\%$ confidence with and without
GW200129\_065458 respectively.
Using four different parametric population models for eccentricity, we argue that the rate of eccentric events is weakly constrained by observations and highly model-dependent.
\end{abstract}
\maketitle

\section{Introduction}
\label{sec:intro}

The catalog of gravitational-wave (GW) sources
\cite{2019PhRvX...9c1040A,2021PhRvX..11b1053A,2024PhRvD.109b2001A,LIGO-O3-O3bcatalog,LIGO-O4a-cbc-catalog_results} 
observed by the (LIGO-Virgo-KAGRA) LVK detectors \cite{LIGOScientific:2014pky,VIRGO:2014yos,10.1093/ptep/ptaa125} continues to grow
as detector sensitivities improve \cite{2015CQGra..32g4001L,2019NatAs...3...35K,2021PTEP.2021eA101A,2025arXiv250818081T,2020LRR....23....3A}.
The larger and more comprehensive catalog increasingly includes new discoveries, with physics or phenomena previously not
confidently apparent from earlier catalogs
\cite{LIGO-O1-BoxingDay,LIGO-GW170817-bns,LIGO-O3-GW190412,LIGO-O3-GW190814,LIGO-O3-GW190521-implications,LIGO-O4-HierarchicalPair-2025,LIGO-O4-GW231123}.
The distinctive properties of these discoveries provide clues into how compact binaries form
\cite{2009LRR....12....2S,Thorne1977,2010CQGra..27k4007M,PSconstraints3-MassDistributionMethods-NearbyUniverse,2016Natur.534..512B,AstroPaper,2017ApJ...846...82Z,2017PhRvL.119y1103V,2022PhR...955....1M,2022ApJ...940..171R,2020ApJ...903L...5R,2021ApJ...921L..43Z,2020FrASS...7...38M}.
As one example, recent observations have confidently identified multiple events each individually consistent with hierarchical compact binary
formation:  very massive black holes with large spin (GW231123 \cite{LIGO-O4-GW231123}), lower-mass asymmetric binaries with large primary black holes (BH) spin (GW241011/GW24110 \cite{LIGO-O4-HierarchicalPair-2025} and GW231118 \cite{2025arXiv250923897L}), and a few events with proposed indications of orbital eccentricity (e.g., GW200105 (NSBH), GW200129, GW200208\_22, GW190701, GW190521,
GW191109, GW190601, and GW190929)
\cite{2019MNRAS.490.5210R,%
2025arXiv251207688S,%
2022NatAs...6..344G,%
2020ApJ...903L...5R,%
2022ApJ...940..171R,
2024ApJ...972...65I,%
2025PhRvD.112j4045G,%
2025ApJ...995...47P,%
2025arXiv250812460J,%
2025arXiv250800179K,%
2025arXiv250315393M,%
2025PhRvD.112f3052R,%
2025arXiv250722862M%
}.
%

More broadly, however, the whole catalog combined can reveal the underlying population, enabling  sharper questions about
compact binary formation channels
\cite{LIGO-O4a-cbc-catalog_methods,2025arxiv250818083T,LIGO-O3-O3a-RP,LIGO-O3-O3bpop,gwastro-PopulationReconstruct-Parametric-Wysocki2018,gwastro-DanielW-PopsynKickPaper2017,gwastro-Davide-PopsynKickPaper2018,gwastro-PopulationReconstruct-Hierarchical-WysockiDoctor2019,popsyn-gwastro-STInterpFinal-Vera2023,gwastro-wd-DelfaveroCosmic-2024,gwastro-agndisk-GayathriPopModels2022,gwastro-agndisk-GayathriPopModels2025}.
For example, the overall distribution of binary spins and trends in primary spin versus mass may provide insight to differentiate between different formation
channels. Some of which might form binaries with preferentially aligned spins and masses generated from isolated binary stellar evolution, and others which invoke more dynamic formation scenarios including hierarchical triples, dense
clusters, or Active Galactic Nuclei (AGN) disks; see, e.g.,  \cite{2025arxiv250818083T,LIGO-O3-O3a-RP,LIGO-O3-O3bpop} and references therein. Many recent investigations point to significant
changes of spin with mass, hinting
at hierarchical formation
\cite{2025arXiv250923897L,2025arXiv250717551L,2025arXiv250915646B,2025arXiv250923897L,2025arXiv250717551L,dcc-Tong-Hierarchical-2025,2025arXiv251025579T,2025PhRvL.134a1401A,2024PhRvL.133e1401L,2022ApJ...928..155T,2022PhRvD.105l3024F,2024arXiv240601679P}.

However, recent investigations suggest that the spin alone may not unambiguously discriminate between these channels:  for example isolated binaries may not always be preferentially
aligned \cite{2024arXiv241203461B,2021PhRvD.103f3007S,2025NewA..12102459L}, while conversely mechanisms in some dynamical environments can produce some spin-orbit alignment
\cite{gwastro-agndisk-McFacts2-2024,2025arXiv250905415K,2025ApJ...983L...9K,2025ApJ...979..237K}, and
triples may lead to entirely different outcomes \cite{2019ApJ...881...41L}.
Orbital eccentricity provides a new GW observable which is inevitably associated with recent strong progenitor dynamics,
given how rapidly GW radiate away
binary eccentricity
\cite{2010CQGra..27k4007M,2021ApJ...921L..43Z,2021ApJ...921L..31R}.
 While
eccentricity too can be imparted in
multiple ways
\cite{2010CQGra..27k4007M,2018PhRvL.120o1101R,2021ApJ...921L..31R,2021ApJ...921L..43Z,2025arXiv250513589S},
this additional parameter provides a
valuable complementary discriminator \cite{2010CQGra..27k4007M,2025ApJ...994L..47S}.
%

Several groups have presented events as candidates for possessing orbital eccentricity \cite{2019MNRAS.490.5210R,%
2025arXiv251207688S,%
2022NatAs...6..344G,%
2020ApJ...903L...5R,%
2022ApJ...940..171R,
2024ApJ...972...65I,%
2025PhRvD.112j4045G,%
2025ApJ...995...47P,%
2025arXiv250812460J,%
2025arXiv250800179K,%
2025arXiv250315393M,%
2025PhRvD.112f3052R,%
2025arXiv250722862M%
}.
So far, these candidates individually have only marginally significant indications of eccentricity, particularly given significant modeling
systematics and strong prior bias against ubiquitous eccentric sources.  
Recently, Gupte et al
\cite{2025PhRvD.112j4045G}  performed a proof-of-concept investigation to ascertain what fraction of
massive compact binary mergers might possess eccentricity.   Using independent eccentric source-parameter inference for
57 massive binary black holes (BBH) (finding 3 candidate eccentric events) and an assumed-known BH mass
and spin 
distribution model (i.e., fixed to the maximum-likelihood LVK O3 result), the authors estimate roughly
$2.4/57$  of
current detections are eccentric, consistent with the number of sources identified as potentially eccentric with their
source-parameter inference. 
Additionally, Singh and
collaborators 
\cite{2025arXiv251207688S} investigated the sensitivity of current GW searches to eccentricity, to ascertain
under what conditions a measurable eccentricity could be identified.   Their analysis found that many of the
commonly-proposed candidates for orbital eccentricity -- notably the same three candidates identified by Gupte and
collaborators (GW200208\_22, GW200129, and GW190701) -- have properties such that orbital eccentricity
could indeed be measured. 

%
In this paper, we perform
a joint analysis of a comprehensive sample of events to ascertain the overall evidence for eccentricity in the full GW population,
fitting for the population including a variable, unknown eccentricity distribution.  For direct comparison with O4a BBH population studies \cite{2025arxiv250818083T}, our sample set includes same 153 confident BBH with false alarm rate (FAR) less than 1 per year, so explicitly excludes the highly significant neutron star black hole (NSBH) candidates, such as GW200105. To draw our conclusions, we perform independent parameter estimation (PE) for all GW candidates using \textsc{SEOBNRv5EHM} model including orbital eccentricity, then reassess the overall population using the \gwk population inference engine.
Building on our prior investigations \cite{2024PhRvD.110f3009Z} with synthetic data to tease out evidence for eccentricity from GW sources, we perform the first measurement of the GW population allowing for flexible mass, spin,
and eccentricity distributions, as well as incorporating events with all mass scales.
Our underlying source-parameter inferences include one source (GW200129\_065458) with a priori evidence for
eccentricity, relative to a uniform prior. However, the evidence for eccentricity in this event depends the specific
choices adopted when performing PE; therefore, we perform the population inference with and without
this event. Similarly, because conclusions about  GW231123\_135430 depend on the choice of waveform, mainly for
redshift, therefore, we also perform the population inference with and without this event to assess its impact on the
inferred redshift distribution.  Unsurprisingly, we therefore effectively find upper limits on the population of eccentric sources,
within the context of a population fit that otherwise qualitatively recovers the same features
previously identified in the original GWTC-4 analysis \cite{2024PhRvD.110f3009Z}.

This paper is organized as follows: \Cref{sec:PE} summarize the eccentric PE and key observations of the posteriors.
\Cref{sec:methods} explains the methods used for the study, such as population likelihood, detection model, and population model for eccentricity. \Cref{sec:results}, present our studies of eccentricity distribution and comparison of the population with circular PE. \Cref{sec:conclude} concludes the key findings and outlines directions for future work. Finally, \Cref{sec:appendix} explain the basic assumptions made to perform the analysis, a table to show the values of PE being used with Bayes factor, and brief discussion on waveform systematics seen on PE level.

\section{Parameter Estimation (PE)}
\label{sec:PE}

\subsection{Gravitational-Wave Observations and Interpretation}

\begin{figure*}[ht]
    \centering
    \includegraphics[width=0.95\textwidth]{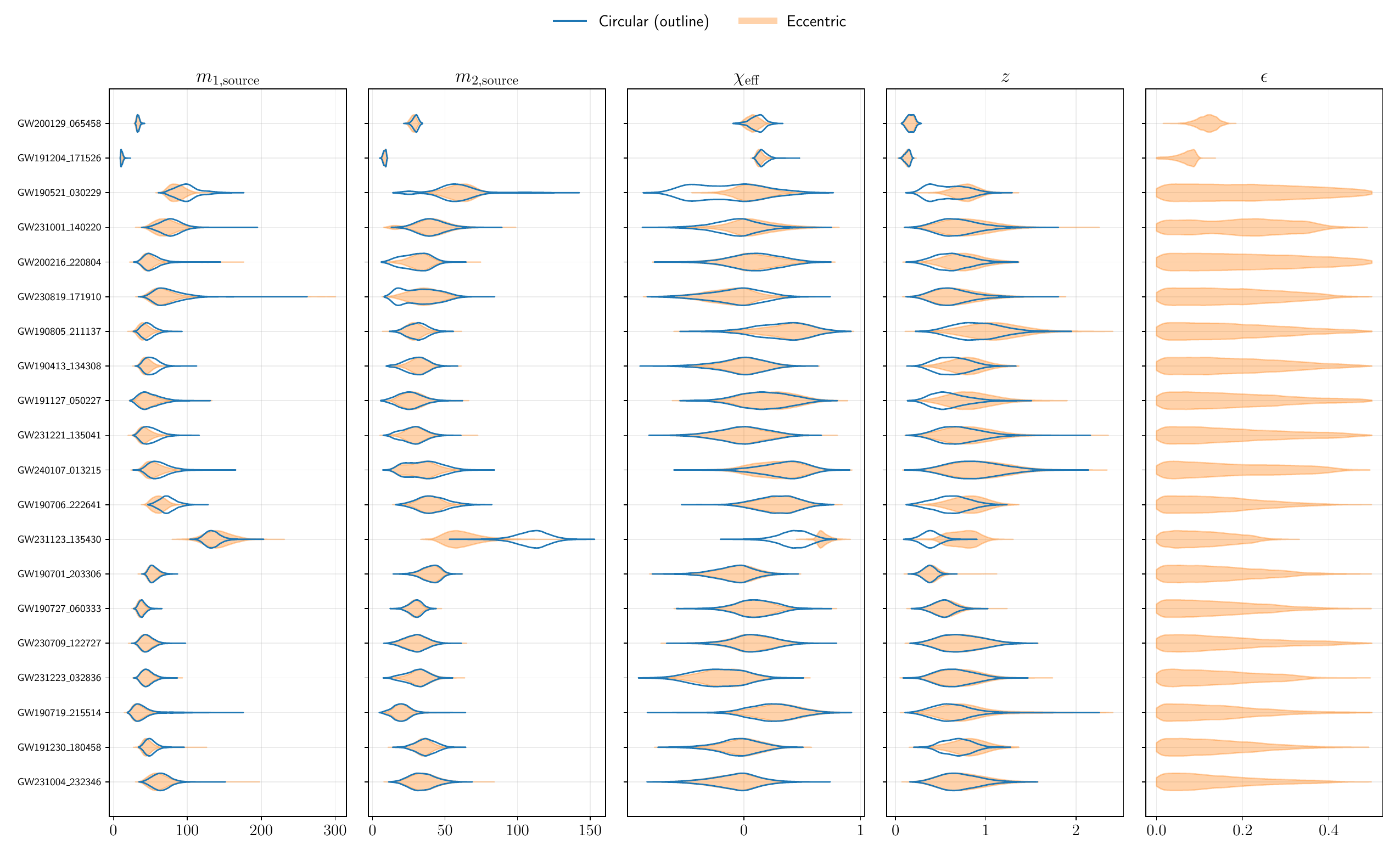} 
    \caption{\textbf{PE Comparison:} The blue color shows the PE with quasi-circular assumption, published under GWTC-4 catalog on zenodo vs orange color shows PE performed with RIFT using eccentric waveform \textsc{SEOBNRv5EHM}. We show only top twenty events from the \Cref{tab:event_summary} where they are sorted based on the bayes factor for eccentricity, detailed in appendix.}
    \label{fig:vilon}
\end{figure*}

We consider all BBH until O4a satisfying a conservative selection criteria
($\rm FAR < 1~yr^{-1}$, minimum over all pipelines) and characterized in either GWTC-2 \cite{LIGO-O3-O3a-catalog}, GWTC-3 \cite{LIGO-O3-O3b-catalog},
GWTC-2.1 \cite{LIGO-O3-O3a_final-catalog}, or GWTC-4 \cite{LIGO-O4a-cbc-catalog_results}. \Cref{tab:event_summary}
enumerates the 153 BBH events
identified by this criteria, which is consistent with the choices adopted in the LVK GWTC-4 population study \cite{2025arxiv250818083T}.
For each candidate event, we employ source-parameter inferences reported by Malagon et al
\cite{2026arXiv260512818M}, performed using the \textsc{SEOBNRv5EHM} waveform model \cite{2025PhRvD.112d4038G}
with the RIFT parameter-estimation
framework \cite{gwastro-PENR-RIFT,gwastro-PENR-RIFT-GPU,gwastro-RIFT-Update,gwastro-RIFT_FinerNet}.
The \textsc{SEOBNRv5EHM} model assumes the binaries have spin angular momenta parallel or antiparallel to the binary orbital
angular momentum, does not have spin precession.  Unless otherwise noted, we employ the nominal waveform-dependent eccentricity for each event defined at the start of the
waveform, which for these analyses is usually 10 Hz; see Malagon et al \cite{2026arXiv260512818M} for details.
Each event analysis provides roughly $8\times 10^4$ independent posterior samples, of which we randomly select 5000 per event for population analysis.
\Cref{fig:vilon} compares selected parameter inferences with eccentricity used in this work to the corresponding parameter inferences previously reported in GWTC catalogs with quasi-circular assumption.  Here and in the catalog, the conclusions derived using an eccentric waveform model are extremely consistent with previously reported results, excepting only a handful of special cases discussed at greater length in \cite{2026arXiv260512818M}.

\Cref{tab:event_summary} summarizes salient properties of these events, as inferred self-consistently with eccentricity.
While most events are consistent with zero eccentricity, several events' marginal likelihoods peak away from zero
eccentricity, indicating consistency with potentially nonzero eccentricity.
To quantify this nominal significance, \Cref{tab:event_summary} includes a fiducial Bayes factor relative to a uniform prior on
eccentricity; see \Cref{sec:appendix} for some details and \cite{2026arXiv260512818M} for extensive discussion. Only one of the events in our catalog show significant evidence for eccentricity, even when adopting this extremely favorable prior.

\begin{figure}
    \centering
    \includegraphics[width=0.45\textwidth]{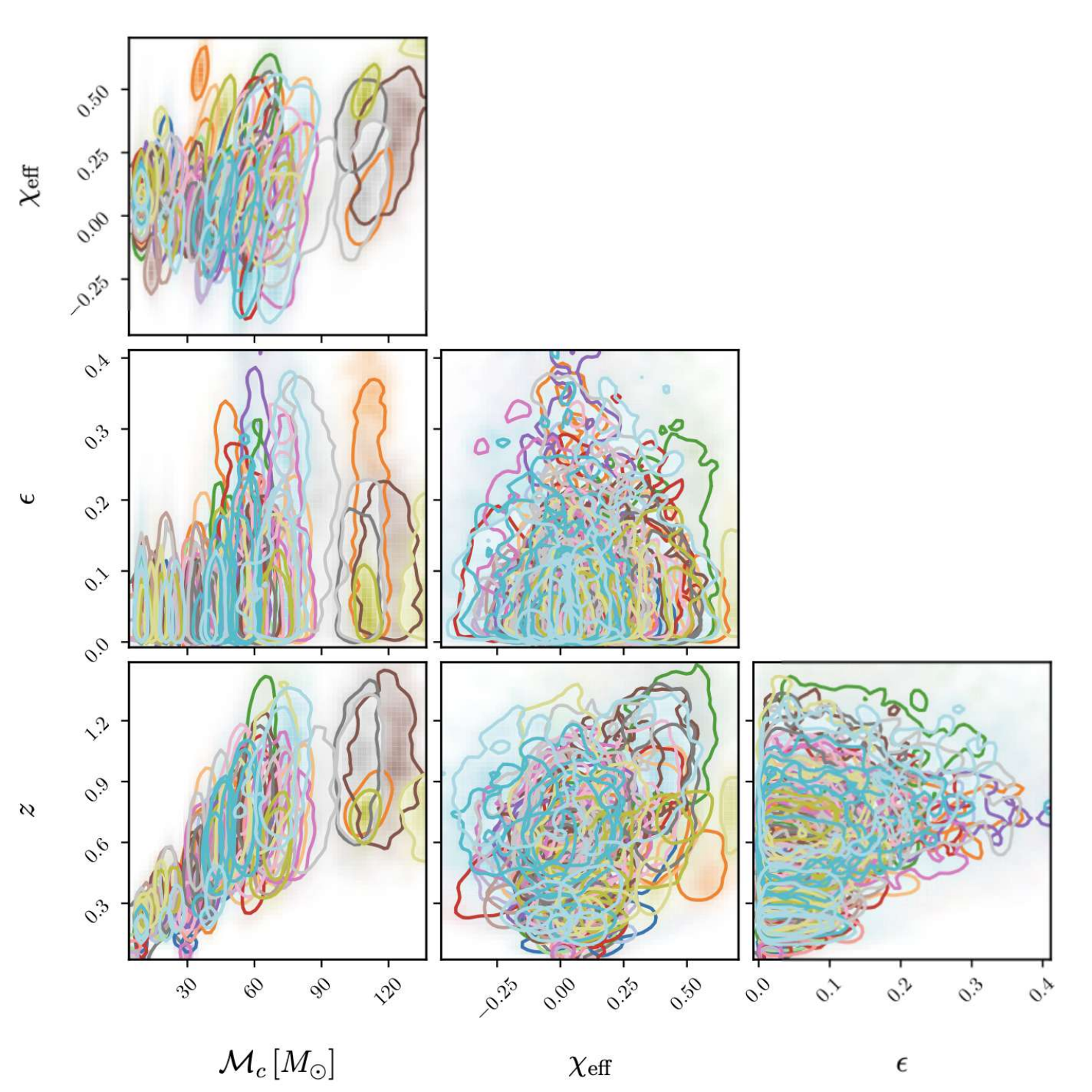} 
    \caption{\textbf{PE Corner Plot:} This plot shows the posterior distributions of the chirp mass $\mathcal{M}_c$, effective spin $\chi_{\mathrm{eff}}$, eccentricity $\epsilon$, and redshift $z$ for a representative eccentric BBH merger. The contours represent the $50\%$ credible intervals for each event, providing insights into the correlations between these parameters.}
    \label{fig:corner_q_chieff_e_z}
\end{figure}

\Cref{fig:corner_q_chieff_e_z} shows a corner plot superimposing the two-dimensional marginal posteriors of all events versus chirp mass ${\cal M}_c$, redshift $z$, inspiral effective spin $\chi_{\rm eff}$ and eccentricity
$\epsilon$, illustrating salient correlations between parameters. First and foremost, the trend between chirp mass and redshift reflects the tendency of louder sources to be accessible farther
away. 
Second, as expected, eccentricity posteriors' extent shows a strong trend versus redshift.  This trend reflects the tendency of farther-away and higher-mass sources to be seen farther
away; both fainter and shorter signals provide fewer opportunities to constrain the source eccentricity, implying wider
and less informative posteriors. Third, the eye is drawn towards what seems to be a trend towards broader posteriors
at larger $\chi_{\rm eff}$.  This correlation can be qualitatively understood as a reflection of the strong nominal trend between $z$ and
$\chi_{\rm eff}$ apparent in detected events, particularly the events at extreme $\chi_{\rm eff}$.

\begin{figure}[t]
    \centering
    \includegraphics[width=0.45\textwidth]{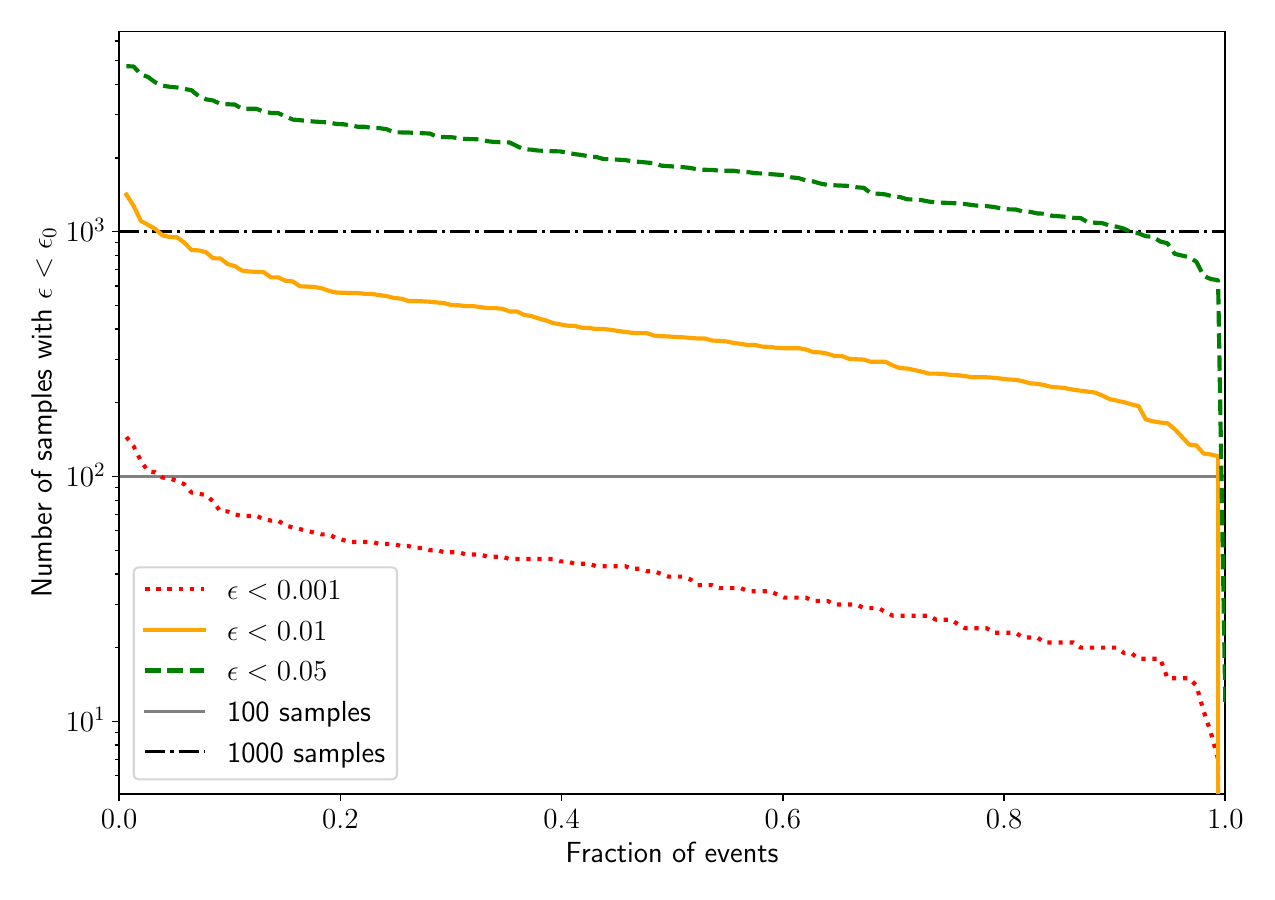} 
    \caption{\textbf{Eccentricity Support:} This CDF shows the low eccentricity support in 153 BBHs whose PE performed with RIFT using \textsc{SEOBNRv5EHM}. This demonstrates that for both $\epsilon=0.01$ and $0.05$ we will have at least 100 independent posterior samples with eccentricity
below that threshold, allowing reliable Monte Carlo estimates for the integrals appearing in hierarchical Bayesian inference.}
    \label{fig:ecc_event_cdf}
\end{figure}

Population inference probing multiple observables can quickly exhaust posterior samples provided for any given event, particularly when the population inference probes narrow population properties in one or more observables. In this work, we want to constrain the spin distribution (narrowly peaked towards small $\chi_{\rm eff}$) and the eccentricity
distribution (narrowly peaked towards zero), while simultaneously constraining multiple and potentially narrow Gaussian features in the mass distribution. To that end, \Cref{fig:ecc_event_cdf} shows, for different eccentricity
thresholds, an inverse cumulative distribution to assess how many independent samples are present above three proposed eccentricity thresholds. Specifically, if $N_\alpha(\epsilon)$ is the number of samples in event $\alpha=1,\ldots,N$ with
eccentricity below $\epsilon$, where $\alpha$ is
indexed so $N_\alpha(\epsilon)$ is monotonically decreasing, then \Cref{fig:ecc_event_cdf} represents $(N_\alpha(\epsilon),\alpha/N)$.  This Figure demonstrates that for both $\epsilon=0.01$ and $0.05$ we will have at least 100 independent posterior samples with eccentricity
below that threshold, allowing reliable Monte Carlo estimates for the integrals appearing in hierarchical Bayesian inference.

\section{Methods}
\label{sec:methods}
\subsection{Review of Hierarchical Bayesian Inference (HBI)}

To infer the BBH population, we adopt the formalism introduced in previous works, referred to as Bayesian parametric models (BPM), implemented in the population inference engine called \gwk \cite{gwkokab2024github,2026PhRvD.113j3003Q}.
Given the likelihood $\ell(\svecz)$ of individual sources and associated reference prior $\pi(\svecz)$, we proceed with a hierarchical Bayesian framework given in \Cref{eq:Bayes_pop} to infer the posterior distribution $p(\pvecz|\data)$ of the BBH population,

\begin{align}
    \label{eq:Bayes_pop}
    \!p\left(\pvecz | \data \right)
     & = \frac{
        \pi(\pvecz)\,
        p(\data | \pvecz)
    }{
        p\left( \data \right)
    },
\end{align}
where $\data=\{{d_j}\}_{j=1}^N$ is the dataset and $d_j$ shows an individual event and N is the total number of events, $p(\pvecz | \data)$ is the posterior distribution of $\pvecz$ given $\data$, $\pi(\pvecz)$ is the population prior on $\pvecz$.  The term $p(\data)$, known as Bayesian evidence, serves as normalization constant and often omitted in sampling-based inference.
Therefore, in practice, we will use the likelihood function $\mathcal{L}(\pvecz)\equiv p(\data | \pvecz)$ to compute the posterior distribution $p(\pvecz | \data) \propto \mathcal{L}(\pvecz)~\pi(\pvecz)$.

To conduct our analysis we have used the inhomogeneous Poisson process \cite{2019MNRAS.486.1086M,2004AIPC..735..195L,PhysRevD.91.023005}

\begin{equation}
    \label{eq:likelihood}
    \mathcal{L}(\pvecz) \propto
    e^{-\mu{(\pvecz)}}
    \prod_{j=1}^N
    \int\ell_j(\svecz) \cdot \comp(\svecz\mid\pvecz)
     \sqrt{ g_{\svecz}}
    d \svecz,
\end{equation}
where exponent $\mu{(\pvecz)}$ is the total expected number of detections under the
given population parametrization $\pvecz$, the complete expression is given in \Cref{eq:mu}. $g_{\svecz}$ is the determinant of the
metric over those coordinates, and  $\comp (\svec | \pvecz)$ is the merger rate density in source frame of reference. For source-parameters, we adopt a usual uniform metric over all intrinsic and extrinsic parameters, such that $\sqrt{g_{\svecz}}d\svecz = T_{\mathrm{obs}} \times dz (1+z)^{-1} (dV_c/dz) \times dm_1 dm_2 \times $  appropriate factors for eccentricity and spin which depend on the coordinate representation adopted for them. The term $\ell_j(\svecz)$ is
the likelihood of individual events and defined as follows,

\begin{equation}
\ell_j(\svecz) \equiv p(d_j|\svecz) \propto \frac{ p(\svecz|d_j)}{\pi(\svecz)}.\label{eq:indi_likelihood}
\end{equation}

The following is the reference prior used for the analysis
\begin{equation}
    \pi(\svecz)=D_L^2(z)\frac{\partial D_L(z)}{\partial z}\times(1+z)^2,
    \label{eq:ref_prior}
\end{equation}

where the factor $(1+z)^2$ converts detector-frame to source-frame masses (primary and seconday), $D_L^2(z)\partial D_L/\partial z$ corresponds to the luminosity-distance prior, further details are given in \cite{2021arXiv210409508C}. In this work, we model only the population distribution of $\chi_{\rm eff}$ and therefore marginalize over all remaining spin degrees of freedom, neglecting the precessing spin parameter $\chi_p$. The reference prior being used on $\chi_{\rm eff}$ is given in Equation 10 of \cite{2021arXiv210409508C}. All integrals appearing explicitly or implicitly in the expressions are computed via Monte Carlo integration, as described in \cite{2026PhRvD.113j3003Q}. Posteriors on hyperparameters $\pvecz$ used in this work are also filtered with the variance of less than 1. See Equation 9, 10 and 11 of \cite{2025PhRvD.111f3043H} for variance of the population likelihood given in \Cref{eq:likelihood}.

\subsubsection{Expected Rate Estimation}
\label{subsection:volume}

The expected number of GW detections can be formulated as an integral over the intrinsic source-parameter space $\svec$ and redshift $z$ modulated by an appropriate selection (weighting) function. The total expected number of detections summing over all populations is given by

\begin{align}
    \label{eq:mu}
    \mu(\pvecz)= \int P_{\mathrm{det}} (\svec;z)\cdot \comp(\svecz\mid\pvecz)\sqrt{ g_{\svecz}}
    d \svecz.
\end{align}

Here $P_{\mathrm{det}}(\svec;z)$ is the detection probability for a source with intrinsic parameters $\svec$ at redshift $z$.

\subsection{Detection Model}
Ideally, to perform eccentric population inference, we need an estimate of search sensitivity with eccentricity. Lacking
comprehensive search results to carefully pin down trends versus eccentricity, we instead rely on previous studies which
suggest low eccentricity $\epsilon<0.4$ have a small effect  impact search
sensitivity for BBH \cite{2010PhRvD..81b4007B,2020ApJ...890....1N,2020PhRvD.102d3005R,2025ApJ...993..215W}, likely less
than 10\% \cite{2024PhRvD.110d4013G}, keeping in mind that some other studies
suggest greater dependence for low-mass NSBH binaries \cite{2025arXiv251210803P}. We therefore use previously-reported search sensitivity estimates neglecting the effects of eccentricity entirely \cite{2025arXiv250818081T,2025PhRvD.112j2001E}.\footnote{We use the semi-analytical sensitivity injections published on zenodo under record-number 16740128 and saved as \url{mixture-semi_o1_o2-real_o3_o4a-cartesian_spins_20250503134659UTC.hdf}.}

We model the astrophysical merger rate density $\rho(\svec, \epsilon, z \mid \pvec_\svec, \pvec_\epsilon, \pvec_z)$ as a joint distribution over source-parameters, where $\epsilon$ denotes the orbital eccentricity, $z$ is the redshift, and $\svec$ collects the remaining intrinsic parameters (e.g., mass, spin). The parameters $\pvec_\epsilon$, $\pvec_z$, and $\pvec_\svec$ represent the corresponding population hyperparameters.

In the present analysis, the sensitivity injections used to estimate the detection
probability do not sample orbital eccentricity. We therefore make the explicit
assumption that the detection probability is independent of eccentricity
\begin{equation}
P_{\rm det}(\svec, \epsilon, z)
\approx
P_{\rm det}(\svec, z).
\label{eq:pdet_independence}
\end{equation}

Under this assumption, eccentricity can be consistently marginalized out of the
selection term,
\begin{equation}
\mu(\pvec_\svec, \pvec_z)
=
\int
P_{\rm det}(\svec, z)\,
\rho(\svec, z \mid \pvec_\svec, \pvec_z)\,
\, d\svec\, dz,
\label{eq:mu_marginalized}
\end{equation}
where
\begin{equation}
\rho(\svec, z \mid \pvec_\svec, \pvec_z)
=
\int_{\epsilon_{\rm min}}^{\epsilon_{\rm max}}
\rho(\svec, \epsilon, z \mid \pvec_\svec, \pvec_\epsilon, \svec_z)\,
d\epsilon.
\end{equation}

The expected number of detections is then estimated using Monte Carlo integration over the
injection distribution,
\begin{equation}
\hat{\mu}(\pvec_\svec, \pvec_z)
=
\sum_{\svec, z \sim \pi_s(\svec, z)}
\frac{
\rho(\svec, z \mid \pvec_\svec, \pvec_z)
}{
\pi_s(\svec, z)
},
\label{eq:mu_estimator}
\end{equation}
where $\pi_s(\svec, z)$ denotes the sampling distribution of the sensitivity injections.

\subsection{Population Model and Priors}
In this study, we used the parametric models published in GWTC-4 population paper \cite{2025arxiv250818083T}. Specifically, for masses, we use \textsc{Broken Power Law + 2 Peaks} model detailed in Equation~B13 and priors on hyperparameters are given in Table~6, for redshift we use \textsc{Power Law} model detailed in Equation B25 and prior is given in Table~7, and \textsc{Skew-Normal Effective Spin} model for the effective spin detailed in Equation~B37 and priors are given in Table~9. We chose the \textsc{Skew-Normal Effective Spin} model for effective spin because it does not assume correlation between $\chi_{\rm eff}$ and $\chi_p$, therefore, we can model $\chi_{\rm eff}$ independent of $\chi_p$, further details can be read in Section 6.3.2 \cite{2025arxiv250818083T} and references therein.

Our population model family in this study extends the above model by a factor $\pi(\epsilon)$ characterizing eccentricity with a two-component
mixture of truncated normal distributions $p(\epsilon|\mu,\sigma)$ defined over interval $[0,0.5]$:
\begin{align}
    \label{eq:ecc_mixture_model}
    \pi(\epsilon)= (1-\zeta_\epsilon) \cdot p(\epsilon \mid 0, \sigma_{\rm circ}) + \zeta_\epsilon \cdot p(\epsilon \mid \mu_\epsilon, \sigma_\epsilon). 
\end{align}
where $\zeta_\epsilon$ shows the branching ratio of eccentric mergers, $\sigma_{\rm circ}$ is the standard deviation of circular binaries, $\mu_\epsilon$ and $\sigma_\epsilon$ are the mean and standard deviation of eccentric binaries. Further to build different models we truncate the both Gaussians with low and high cut offs, \Cref{tab:model_parameters} provides the all values and priors being used for each parameter.
The low-eccentricity population with weight $(1-\zeta_\epsilon)$ has a mean value fixed to zero and a flexible but small
width, with prior bounds chosen to allow sufficiently many samples to resolve any population integrals, as illustrated
in \Cref{fig:ecc_event_cdf}.
The high-eccentricity population with weight $\zeta_\epsilon$ has two independent parameters, allowing a flexible mean value and width, such that the mean value must be significantly different than zero and thus not degenerate with the
low-eccentricity population.
This two-component model is motivated both by the present catalog and by astrophysical modeling, which suggests multiple
formation channels.

To assess how strongly our results depend on prior assumptions, we perform our analyses using four different priors
settings, \Cref{tab:model_parameters} presents these choices. \textsc{Nonoverlapping Mixture} ,
our fiducial model, creates a mixture using two nonoverlapping Gaussians: the low-eccentricity component is exclusively responsible for binaries with $\epsilon \in [0,0.04]$ and the high-eccentricity component is exclusively responsible
for binaries with $\epsilon\in [0.04,0.5]$.  Within each eccentricity interval, the model characterizes the eccentricity distribution with a truncated Gaussian. The low-eccentricity component has an unknown but small
 $\sigma_{\rm circ}$; the high-eccentricity component has an unknown mean and variance, such that the mean must lie within our
 target interval.  \textsc{High Eccentricity Truncated} changes \textsc{Nonoverlapping Mixture} , eliminating the low-eccentricity component by setting $\zeta_\epsilon=1$; to
 simplify sampling, we additionally fix the parameter associated with the (irrelevant) low-eccentricity component.
 \textsc{Overlapping Mixture} model generalizes \textsc{Nonoverlapping Mixture}  by allowing for overlapping Gaussians.  Except for the change in truncation, this model has the same parameter priors as \textsc{Nonoverlapping Mixture}.   Finally, \textsc{Low Eccentricity Truncated} model eliminates the low-eccentricity component by setting $\zeta_\epsilon=1$ in \textsc{Nonoverlapping Mixture} model.  This \textsc{Low Eccentricity Truncated} model allows eccentricities below $0.04$, while \textsc{High Eccentricity Truncated} model requires all events have eccentricity above $0.04$. 

\begin{table*}
\centering
\renewcommand{\arraystretch}{1.2}
\begin{tabular}{c|ccccccccc}
\hline\hline
Model 
& $\rm low_{cir}$ 
& $\rm high_{cir}$ 
& $\rm \mu_{cir}$ 
& $\rm \sigma_{cir}$ 
& $\zeta_\epsilon$ 
& $\rm low_\epsilon$ 
& $\rm high_\epsilon$ 
& $\mu_\epsilon$ 
& $\sigma_\epsilon$ \\
\hline
\textsc{Nonoverlapping Mixture} 
& 0 & 0.04 & 0 & $U(0,0.04)$ & $U(0,1)$ & 0.04 & 0.5 & $U(0.04,0.5)$ & $U(0.04,0.5)$ \\

\textsc{High Eccentricity Truncated }
& 0 & 0.04 & 0 & 1 & 1 & 0.04 & 0.5 & $U(0.04,0.5)$ & $U(0.04,0.5)$ \\

\textsc{Overlapping Mixture} 
& 0 & 0.5 & 0 & $U(0,0.04)$ & $U(0,1)$ & 0 & 0.5 & $U(0.04,0.5)$ & $U(0.04,0.5)$ \\

Low Eccentricity Truncated
& 0 & 0.5 & 0 & 1 & 1 & 0 & 0.5 & $U(0,0.5)$ & $U(0,0.5)$ \\
\hline\hline
\end{tabular}

\caption{\textbf{Mixture Eccentricity Model:} Hyperparameter choices for the eccentricity distribution across the four population models. The fixed, lower and upper truncation values and prior ranges are shown for each parameter of four model choices.}
\label{tab:model_parameters}
\end{table*}

\section{Results}\label{sec:results}

\subsection{Overview and Dependence on Model Assumptions}

We bound the branching ratio for eccentric events to be below $0.051890$, $0.046016$, $0.022011$, and $0.024412$ at $90\%$ confidence with our \textsc{Nonoverlapping Mixture}, \textsc{Overlapping Mixture} \textsc{Nonoverlapping Mixture} (without GW200129\_065458), and \textsc{Overlapping Mixture} (without GW200129\_065458) eccentricity models respectively. We find that conclusions about the prevalence of eccentricity in the population depend somewhat on the assumptions adopted.  For example, both fiducial \textsc{Nonoverlapping Mixture} and alternative \textsc{Overlapping Mixture} conclude that most events are consistent with nearly-zero eccentricity; that the properties of the nearly-eccentric population are extremely poorly
constrained, with $\mu_\epsilon,\sigma_\epsilon$ almost uninformed relative to our priors; and that the quasi-circular
population's width is marginally informed by our observations.  We emphasize that the lower bound on $\sigma_{\rm circ}$
is almost certainly due to finite sample size, see \Cref{fig:ecc_event_cdf}, rather than being data driven.
The single-component truncated models also arrives at qualitatively similar conclusions, though expressed in a different functional representation, as we will see later using a posterior predictive distribution (PPD).
By contrast, the extreme \textsc{Overlapping Mixture} model which requires all binaries have nonzero eccentricity strongly favors Gaussians
tightly concentrated near the (arbitrary) lower bounds we adopted. 

\begin{figure}
    \centering
    \includegraphics[width=0.48\textwidth]{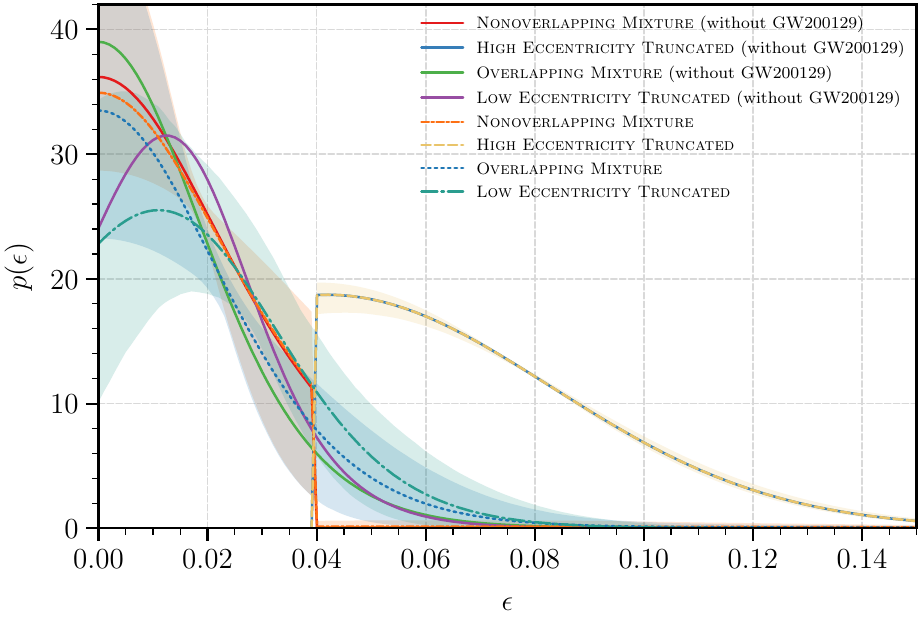} 
    \caption{\textbf{Mixture Eccentricity Model:} This shows a comparison of four models: \textsc{Nonoverlapping Mixture}, \textsc{Overlapping Mixture}, \textsc{High Eccentricity Truncated}, and \textsc{Low Eccentricity Truncated} with and without GW200129\_065458.}
    \label{fig:ecc_ppd}
\end{figure}
\begin{figure}
    \centering
    \includegraphics[width=0.48\textwidth]{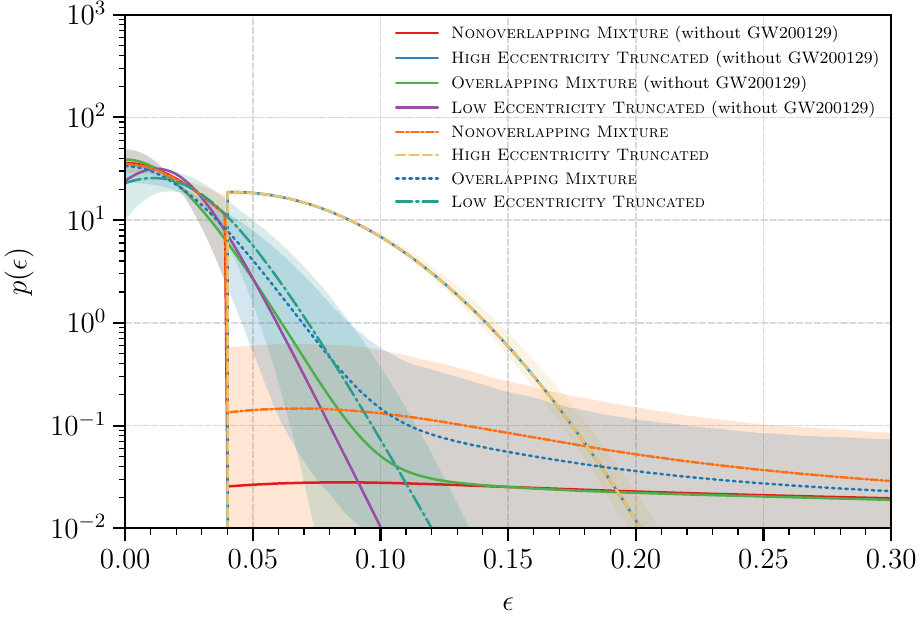} 
    \caption{\textbf{Mixture Eccentricity Model:} This shows the comparison in log scale for four models: \textsc{Nonoverlapping Mixture}, \textsc{Overlapping Mixture}, \textsc{High Eccentricity Truncated}, and \textsc{Low Eccentricity Truncated} with and without GW200129\_065458.}
    \label{fig:ecc_ppd_log}
\end{figure}
\begin{figure}
    \centering
    \includegraphics[width=0.45\textwidth]{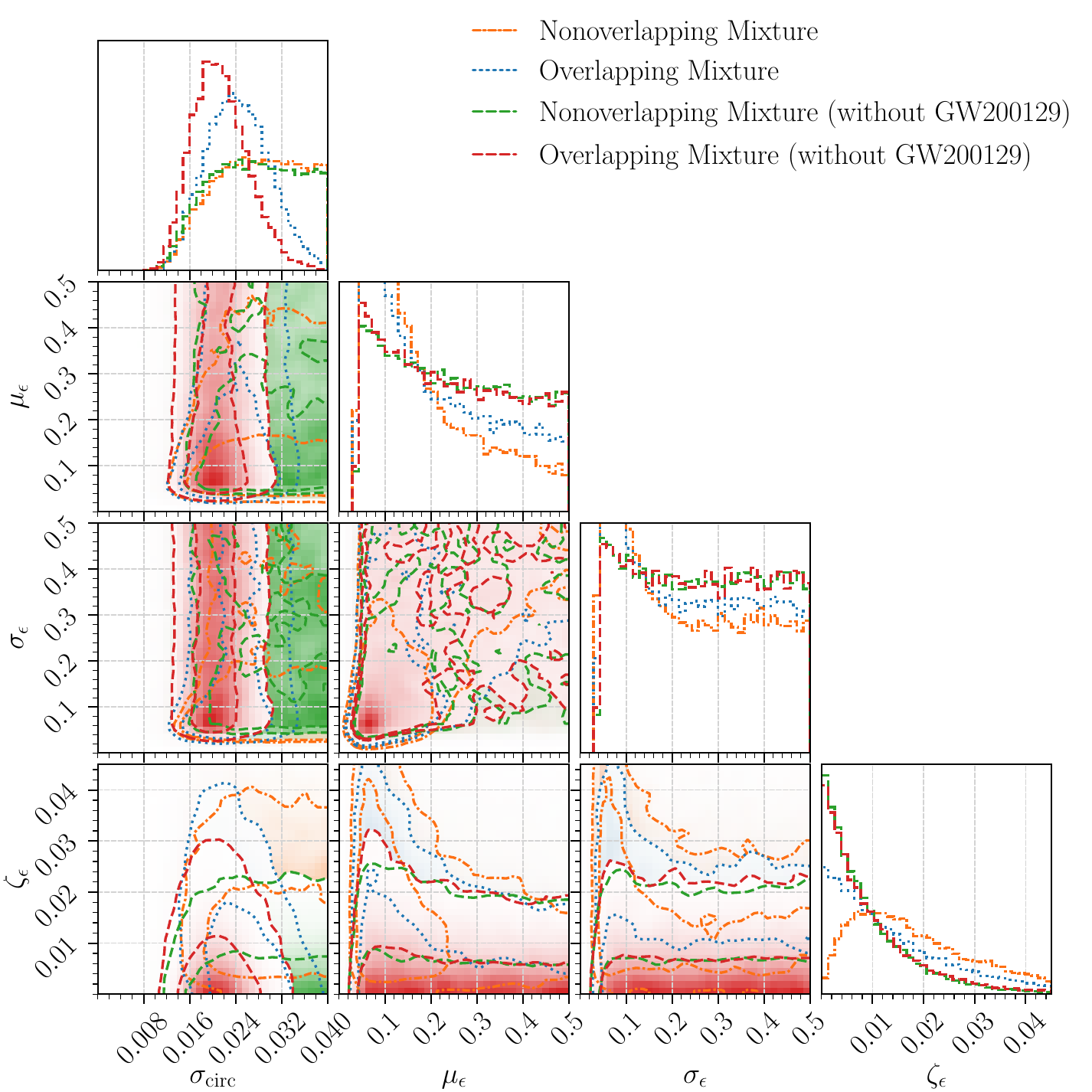} 
    \caption{\textbf{Mixture Eccentricity Model:} This shows the hyperparameters recovery of Nonoverlapping and \textsc{Overlapping Mixture} model for
    eccentricity given in \Cref{eq:ecc_mixture_model}, and detailed in \Cref{tab:model_parameters}. Finite sample size introduces an artificial 
    recovered lower bound on $\sigma_{\rm circ}$; see \Cref{fig:ecc_event_cdf}. We bound the branching ratio for eccentric events to be below $0.051890$, $0.046016$, $0.022011$, and $0.024412$ at $90\%$ confidence with our \textsc{Nonoverlapping Mixture}, \textsc{Overlapping Mixture} \textsc{Nonoverlapping Mixture} (without GW200129\_065458), and \textsc{Overlapping Mixture} (without GW200129\_065458) eccentricity models respectively.}
    \label{fig:corner_mix_ecc}
\end{figure}

\Cref{fig:ecc_ppd} and \Cref{fig:ecc_ppd_log} show the mean and confidence interval for binary eccentricity derived from 
each of four models adopted in this work. These figures illustrate how our strong prior modeling assumptions can lead to different conclusions about the prevalence of low ($\epsilon<0.04$) and more substantial ($\epsilon>0.04$) eccentricity,
despite good qualitative agreement between many of the models locally and even several pairs of the models over a
substantial range (i.e., \textsc{Overlapping Mixture} and \textsc{Low Eccentricity Truncated} at $\epsilon<0.1$, or \textsc{Nonoverlapping Mixture} and \textsc{High Eccentricity Truncated }at $\epsilon>0.15$). \Cref{fig:corner_mix_ecc} show the results of our population inference for eccentric population parameters, for all four
model variations presented above.

\subsection{Interpreting Constraints on eccentricity}

For models \textsc{Nonoverlapping Mixture}, \textsc{Overlapping Mixture}, and \textsc{Low Eccentricity Truncated}, as expected since no individual event shows strong evidence for
eccentricity,  we limit the fraction of potentially highly eccentric events $\zeta_{\epsilon}$
to be below $\simeq O(0.5)/\sqrt{N}$ where $N=153$.  For similar reasons, for the two-component models, we do not draw any conclusions about the properties of the highly-eccentric population: both $\mu_\epsilon$ and $\sigma_\epsilon$ are almost uninformed by the
data relative to their priors.
For the low-eccentricity component, we find an upper bound on its width $\sigma_{\rm circ}$,  also close to the characteristic $0.5/\sqrt{N}$
value expected given our sample size.  While our population inference nominally also bounds this quantity below, this feature almost certainly reflects the collapse of available samples for inference illustrated in \Cref{fig:ecc_event_cdf}.

\Cref{fig:ecc_ppd} and \Cref{fig:ecc_ppd_log} show 
that most inferred eccentricities are small, and that we cannot meaningfully rule out relatively rare high-eccentricity events with the current catalog, without making very strong assumptions (i.e., as in model \textsc{High Eccentricity Truncated}). 
Indeed, these figures shows that the flexible mixture models \textsc{Nonoverlapping} and \textsc{Overlapping} both predict an eccentricity distribution which is nearly
uniform,  by construction normalized so that constant value is roughly $2 \zeta_\epsilon$.  As discussed above, because
observations provide few confidently eccentric events to constrain the distribution of eccentric binaries, the inferred
population is necessarily uninformed and their structure interprets the absence of evidence as a uniform posterior.  By contrast, the much more aggressive \textsc{Low Eccentricity Truncated} and \textsc{High Eccentricity Truncated} models which allow for only one globally Gaussian
population, by construction arrive at prior-dominated conclusions. By design, they each suggest that some eccentricity range is rarely populated.

As discussed at greater length in Malagon et al \cite{2026arXiv260512818M}, the interpretation of GW200129\_065458
depends somewhat sensitively on the analysis choices adopted during parameter inference. To assess the impact of this
event on our conclusions about the population,  we perform the population inference with and without
GW200129\_065458. \Cref{fig:ecc_ppd} and \Cref{fig:ecc_ppd_log} compare results derived with and without this event,
showing comparable analyses produce comparable results -- even though this event provides individually the strongest
indications of eccentricity, its exclusion does not substantially change our conclusions about the population. The sole
exception is the branching ratio in the nonoverlapping mixture case, where the probability density is very small for
$\zeta_\epsilon$ if GW200129 is included but large if excluded, as expected: one event with overwhelming evidence for eccentricity
would rule out $\zeta_\epsilon=0$ exactly, but be consistent with a small but nonzero branching ratio.  In this analysis, we bound the branching ratio for eccentric events to be below $0.051890$, $0.046016$, $0.022011$, and $0.024412$ at $90\%$ confidence with \textsc{Nonoverlapping Mixture}, \textsc{Overlapping Mixture}, \textsc{Nonoverlapping Mixture} (without GW200129\_065458), and \textsc{Overlapping Mixture} (without GW200129\_065458) eccentricity models respectively.

\subsection{Eccentric vs Circular Population Inference}
\begin{figure*}
    \centering
    \includegraphics[width=0.95\textwidth]{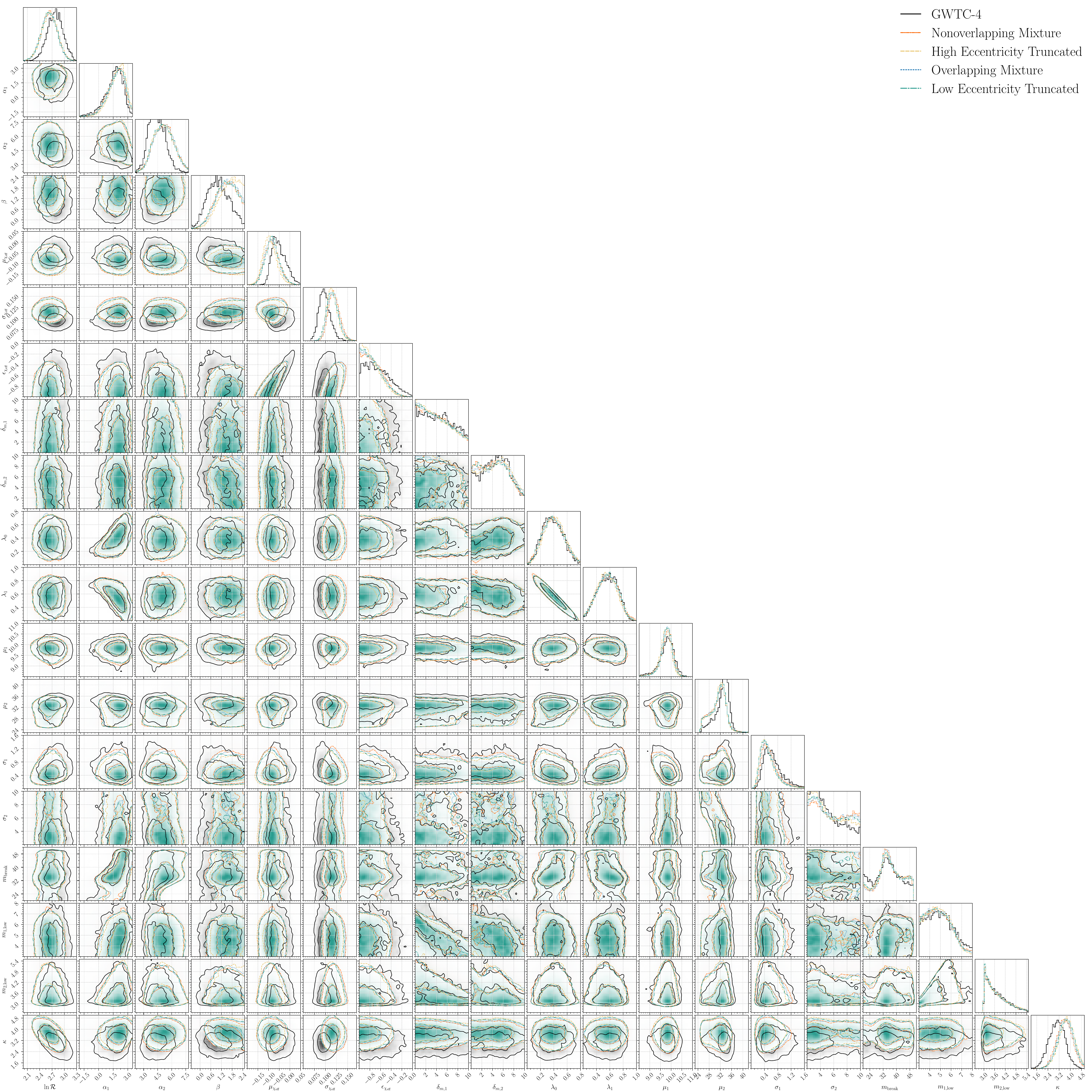} 
    \caption{\textbf{Infered Hyperparameters of Population Models:} This shows the comparison of GWTC-4 catalog published results assuming quasi-circular binaries vs RIFT PE using \textsc{SEOBNRv5EHM} waveform which includes eccentricity.}
    \label{fig:corner_pop_overplot}
\end{figure*}

\Cref{fig:corner_pop_overplot} shows the inferred model hyperparameters obtained in our fiducial analyses (using
eccentric source-parameter inference), compared a reanalysis of GWTC-4 using a comparable model but neglecting
eccentricity (using LVK quasi-circular source-parameter inference).\footnote{We use the dataset published on zenodo under record-number 16911563, and saved as \url{BBHSpin_EpsSkewNormalChiEff.h5}.} For almost every one- and two-dimensional marginal distribution, the posterior
distributions are very similar or identical.  Very few directly interpretable phenomenological parameters show  modest differences:
notably, the mean and width of the spin distribution ($\mu_{\chi_{\rm eff}},\sigma_{\chi_{\rm eff}}$), redshift index $\kappa$, and slope of second powerlaw $\alpha_2$. Other small
differences are associated with inferred smoothing parameters; the smallest allowed secondary mass; and the width of
Gaussian features in the modeled distribution of primary masses. 
The extremely close similarity between these two results suggests that our population inferences are robust.
Indeed, we see very close agreement between these two analyses despite using independent source-parameter inferences, including different
physics (i.e., allowing eccentricity but forbidding precession).

\begin{figure}
    \centering
    \includegraphics[width=0.45\textwidth]{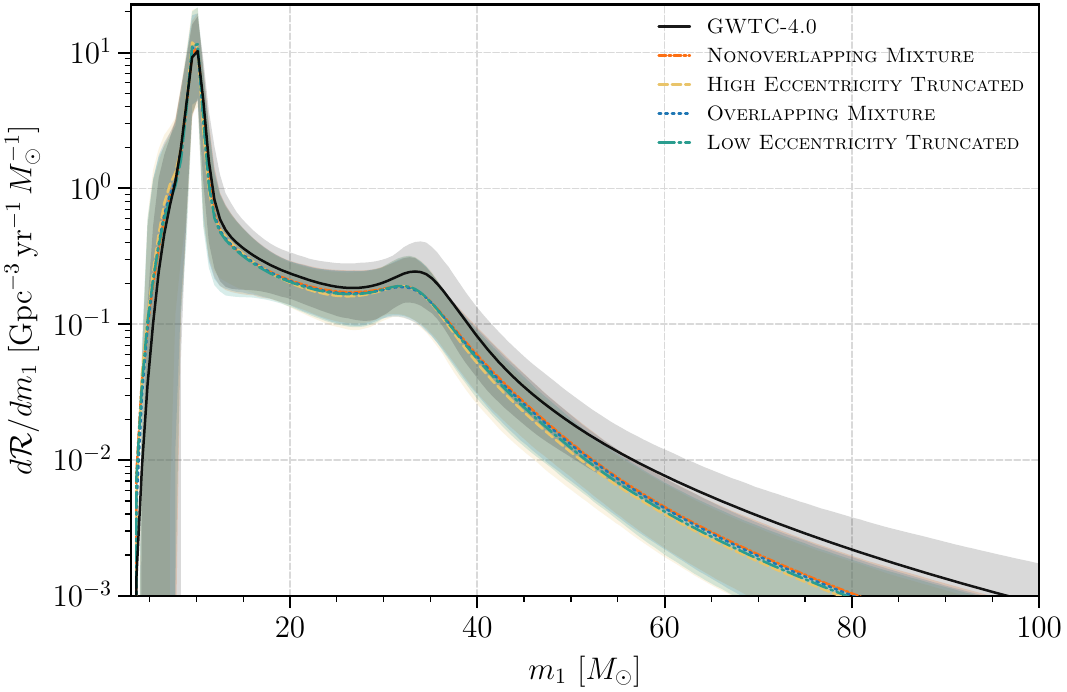} 
    \caption{\textbf{Primary Mass:} This shows the comparison of GWTC-4 catalog published results assuming quasi-circular binaries vs RIFT PE using \textsc{SEOBNRv5EHM} waveform which also models eccentricity. All the four models shows the consistent results with GWTC-4 but slope $\alpha_2$ of second powerlaw is higher than GWTC-4 population analysis with quasi-circular assumption.}
    \label{fig:primary_mass}
\end{figure}

\begin{figure}
    \centering
    \includegraphics[width=0.45\textwidth]{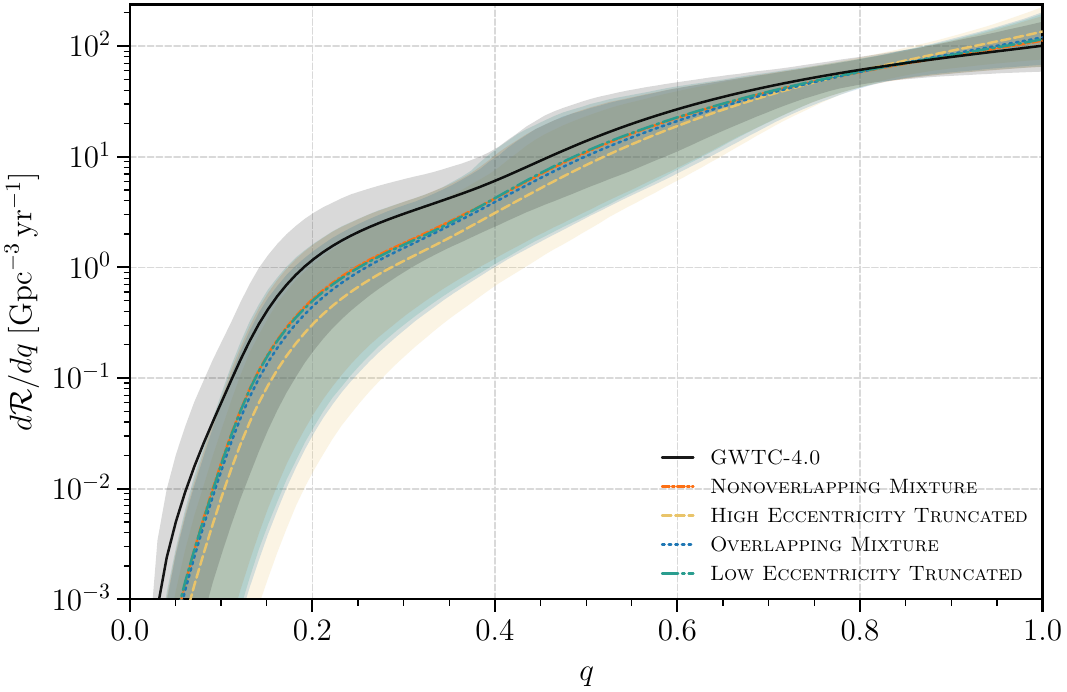} 
    \caption{\textbf{Mass Ratio:} This shows the comparison of GWTC-4 catalog published results assuming quasi-circular binaries vs RIFT PE using \textsc{SEOBNRv5EHM} waveform which also models eccentricity. All the four models shows the consistent results but shows minor differences with GWTC-4. Those differences arise from the PE being used, they show slightly different results of primary and secondary masses for few events which may affects the mass-ratio. The details of the PE are given \cite{2026arXiv260512818M}.}
    \label{fig:mass_ratio}
\end{figure}

\Cref{fig:primary_mass} and \Cref{fig:mass_ratio} compares the PPDs for primary mass and mass ratio for the two approaches to each
other and to previously-published results from GWTC-4.  As expected given major hyperparameter agreements, and minor disagreements,
the all of our results agree well with one another. For the primary mass, our four analyses also agree well with the published GWTC-4 results before the $33 M_\odot$ peak. In other words, the slope ($\alpha_1$) of first powerlaw before before the $33 M_\odot$ extremely well, but we find the higher slope ($\alpha_2$) for the high mass powerlaw after the  $33M_\odot$. For binary mass ratio $q$, our estimates differ slightly from the
previously published GWTC-4 result.  Keeping in mind our parameter inferences are completely independent, with different
physics, and occasionally recover slightly different mass ratios for individual events (\Cref{fig:vilon}, and \Cref{fig:sys}), small
differences between our analysis and the GWTC-4 results are expected. Overall, our eccentric population analysis and PE are producing similar
answers to the previously-published analysis.

\begin{figure}
    \centering
    \includegraphics[width=0.45\textwidth]{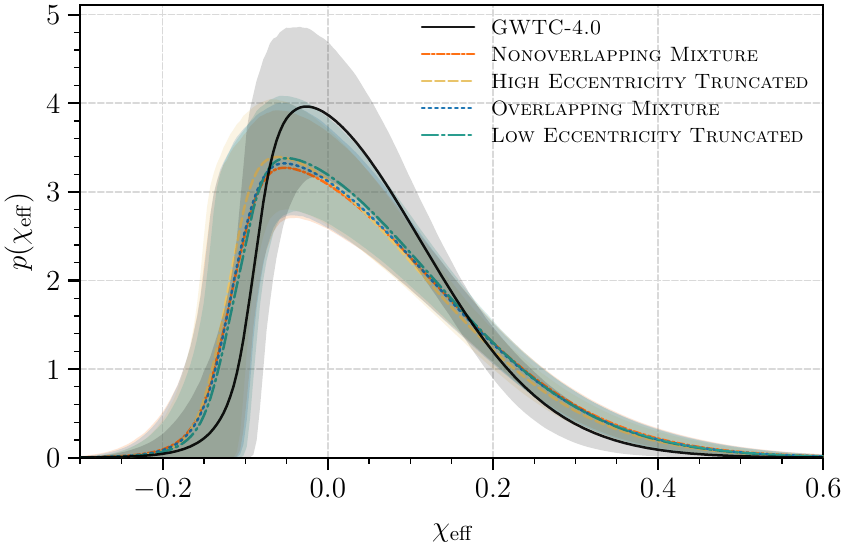} 
    \caption{\textbf{Effective Spin:} This shows the comparison of GWTC-4 catalog published results assuming quasi-circular binaries vs RIFT PE using \textsc{SEOBNRv5EHM} waveform which also models eccentricity. All the four models shows the consistent results with each other but slightly off from the GWTC-4. The reason of different results may coming from eccentric PE, because, \textsc{SEOBNRv5EHM} only allow the aligned or anti-aligned spin and ignore the spin precession. However  quasi-circular PE and selection effects have full spin with precession.}
    \label{fig:chi_eff}
\end{figure}

\Cref{fig:chi_eff} compares the mean for $\chi_{\rm eff}$ for the four models and to previously
published results from GWTC-4. Despite using completely independent source-parameter inferences, our conclusions
about the spin distributions are in good agreement with previously published conclusions.  As above, our conclusions do not depend notably on our choice of eccentric population model. Our differences between the GWTC-4 results may be traced to small but notable differences in the inferred effective spin for some events using the \textsc{SEOBNRv5EHM} waveform model.

\begin{figure}
    \centering
    \includegraphics[width=0.45\textwidth]{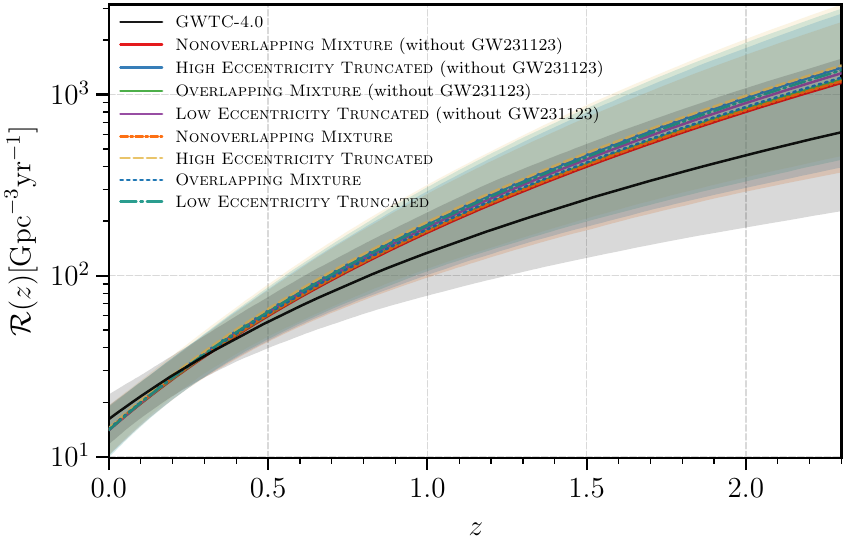} 
    \caption{\textbf{Redshift:} This shows the comparison of GWTC-4 catalog published results assuming quasi-circular binaries vs RIFT PE using \textsc{SEOBNRv5EHM} waveform which also models eccentricity. All the four models shows the consistent results with each other and GWTC-4 with and without GW231123\_135430.}
    \label{fig:redshift}
\end{figure}

\Cref{fig:redshift} compares the PPDs for redshift for the four model choices and to GWTC-4 results. All of our model
    choices are very consistent with each other, including analyses that omit GW231123..  Relative to quasi-circular GWTC-4 results, we recover a 
    slightly more rapid evolution of rate with redshift (i.e., a higher value of the redshift evolution parameter
    $\kappa$), albeit within the confidence interval of previously published results.   We ascribe the modest
    differences seen to systematic differences between the physics and waveforms adopted in our analysis (eccentric but
    non-precessing with SEOB)
    and the physics and waveforms adopted in the GWTC-4 analysis (quasi-circular precessing events with heterogeneous
    waveform modeling): different parameter inference choices produce slightly different results for intrinsic and
    extrinsic parameters including redshift, which can directly impact recovered population parameters.   In
    Appendix \ref{sec:appendix} we concretely demonstrate parameter differences between eccentric and quasi-circular
    analyses for some of the events used, for purposes of illustration.

\section{Conclusion}
\label{sec:conclude}

In this paper, we reanalyzed the significant 153 BBH events ($\rm FAR < 1~yr^{-1}$) from GWTC-4 catalog, using
source-parameter inferences and population models which allowed for orbital eccentricity. Finding significant evidence
for eccentricity only in one event but not in the population overall. We furthermore show the inferred distribution of
highly eccentric objects depends strongly on prior modeling assumptions about the population.   By contrast, the
inclusion or omission of the single event with the largest statistical evidence for eccentricity to account for
substantial systematics (GW200129\_065458)  does not necessarily materially change our
conclusions about the population.  For some sets of assumptions about the population, this source would unambiguously
indicate a nonzero branching ratio for significantly eccentric events; for other frameworks, however, this event
provides minimal new information.

Other than eccentricity, our analysis largely replicates the headline conclusions presented in an analysis of the GWTC-4 population, despite using completely different waveform models, source-parameter inference, and underlying source-population inference engine called \gwk \cite{gwkokab2024github,2026PhRvD.113j3003Q}. Our analysis is the very first to fit all features of the BBH population while accounting for orbital eccentricity.
Our work demonstrates that orbital eccentricity can be efficiently incorporated into conventional analyses of source-parameters and populations, producing robustly similar results, and provides a prototype and benchmark for all such future studies.

Our study has a few limitations, owing to the limited catalog, waveform modeling, searches with quasi-circular assumption, sensitivity injections omitting eccentricity and population modeling used.
For example, in our study of BBH, we omitted the low-mass NSBH candidate GW200105, which has the most tantalizing evidence for eccentricity to date.
Too, our source-parameter and population models allow for non-precessing binaries, even though spin precession and orbital eccentricity can be difficult to differentiate in short BBH signals \cite{2023MNRAS.519.5352R,2026arXiv260102260T,2024PhRvD.109d3037D,PhysRevLett.126.201101}.  For
example, analyses of GW231123  provide no evidence for orbital eccentricity in models that include both orbital eccentricity and precession, but misleadingly positive evidence when compared to models which
omit spin precession \cite{2025arXiv251220060J}. Future work should apply a more generic waveform and population model.
Finally, relative to the roughly log-uniform eccentricity distributions often predicted astrophysically for
eccentricity, the Gaussian population models adopted here seem very optimistic as regards the prospects for measurable
eccentricity.  Given substantial ongoing reassessment of the astrophysical formation channels and outcomes for
BBH, including subpopulations with observationally accessible eccentricity, detailed comparison against
astrophysical formation predictions should always be complemented by purely phenomenological approaches which allow the
data to speak for itself.  Our simple eccentricity model is both sufficiently small and conservative while being meaningfully be constrained by the data.
As more events with evidence for eccentricity accumulate, a more flexible model allowing for multiple populations and correlations between eccentricity and other parameters could be warranted.

Our results contrast with previous claims suggesting several events exhibited substantial eccentricity and thus that a substantial subpopulation of eccentric events was required by current observations. Malagon et
al \cite{2026arXiv260512818M} describes how our source-parameter inferences differ from previously reported results with this and other waveforms.  Our conclusions follow from the parameter inferences provided by that study, only one event provides strong
evidence for significant eccentricity with the \textsc{SEOBNRv5EHM} waveform model.


\section*{Acknowledgements}
This material is based upon work supported by the NSF's LIGO Laboratory, a major facility fully funded by the National Science Foundation. The authors acknowledge the computational resources provided by the LIGO Laboratory's CIT cluster, which is supported by National Science Foundation Grants PHY-0757058 and PHY0823459. ROS acknowledges support from NSF Grant No. AST-1909534, NSF Grant No. PHY-2012057, and the Simons Foundation.

\section{Appendix}\label{sec:appendix}
\subsection{Bayes Factor}\label{subsec:bayes_factor}
We have computed the Bayes factor to find the evidence against the quasi-circular orbit in each event. Specifically, we compute it by comparison of eccentric hypothesis $(\mathcal{H}_1)$ against the quasi-circular limit $(\mathcal{H}_0)$ using the Savage-Dickey density ratio. The Bayes factor in favor of quasi-circular events for a uniform prior on eccentricity $[0,0.5]$ is computed as follows
\begin{equation}
    \mathrm{BF}_{01} = \frac{p(\epsilon=0 \mid \data)}{p(\epsilon=0)} = 0.5 p(\epsilon=0 \mid \data),
    \label{eq:bayes_factor}
\end{equation}

where $p(\epsilon=0 \mid \data)$ is the posterior density evaluated at $\epsilon=0$. We calculate this with boundary corrected kernel density estimator (KDE) constructed by reflection about $\epsilon=0$. We use the Silverman's rule-of-thumb for KDE bandwidth for each event, and the reflected estimator integral is unity over $\epsilon=[0,0.5]$. This schematic provides the Bayes factor in favor of eccentricity as follows

\begin{equation}
    \ln (\mathrm{BF}_{10})= -\ln (\mathrm{BF}_{01}) = -\ln [0.5 p(\epsilon=0 \mid \data)]\label{eqn:b_factor}.
\end{equation}

The positive values of $\ln (\mathrm{BF}_{10})$ show increasing evidence against the perfectly quasi-circular orbits ($\epsilon\neq0$), and negative values show more support for $\epsilon=0$. The smallest negative number $\ln (\mathrm{BF}_{10})$ in \Cref{tab:event_summary} shows the strongest support for quasi-circular case ($\epsilon=0$). We do not find strong evidence for eccentricity for any event, the calculations for each event are shown in \Cref{tab:event_summary}.

\subsection{Waveform Systematics}\label{subsec:systematics}

\begin{figure*}[ht]
    \centering
    \includegraphics[width=\textwidth]{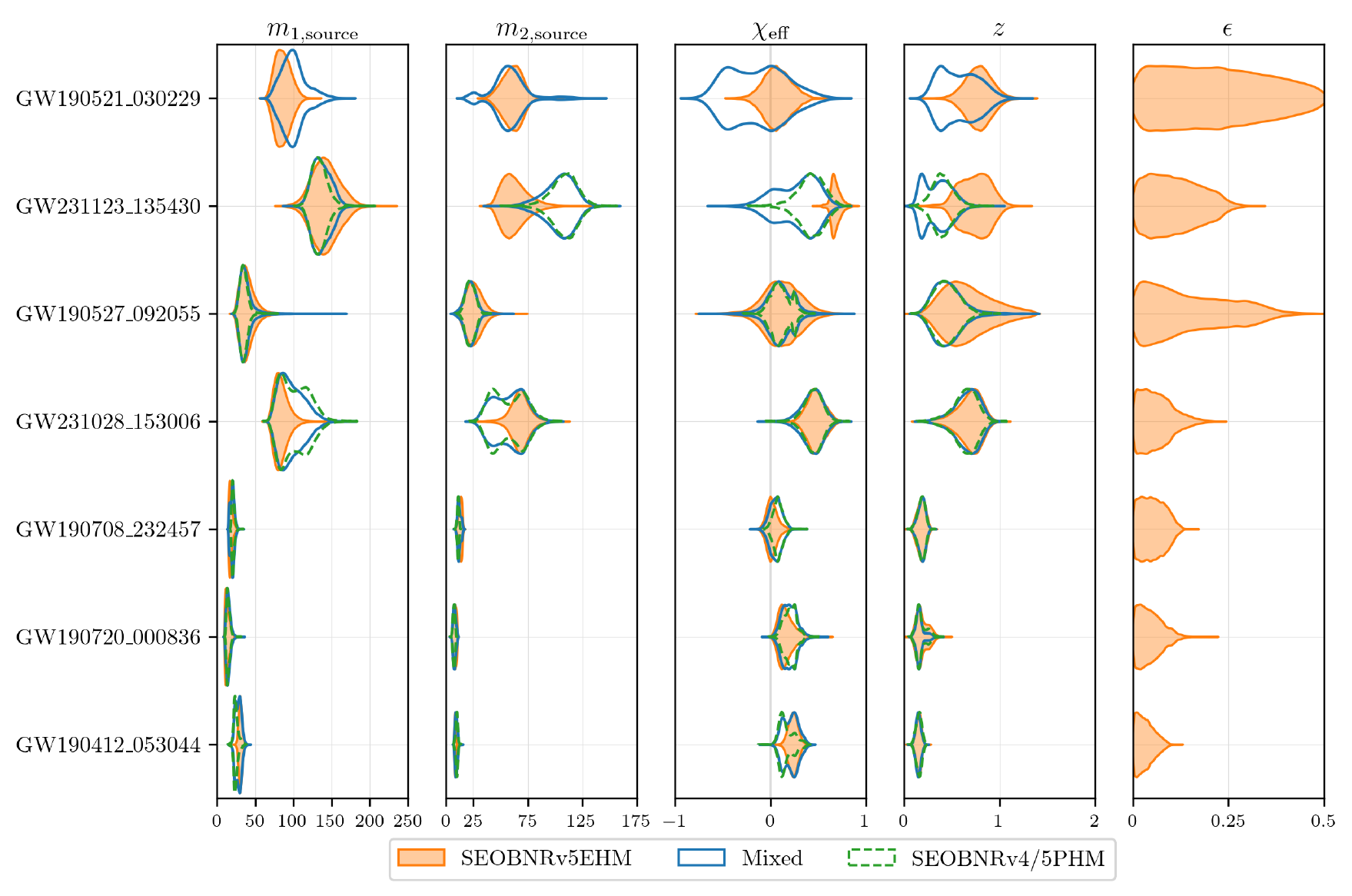} 
    \caption{\textbf{Waveform systematics:} Marginal probability distributions for the source-frame component masses, effective spin, redshift, and eccentricity for events with waveform systematics. The SEOBNRv5EHM result is compared with the mixed and precessing results from public GWTC-2.1 and GWTC-4 data. The events are ordered by descending $\ln$BF$_{10}$.}
    \label{fig:sys}
\end{figure*}

There is a select sample events with notable waveform systematics: GW190412\_053044, GW190521\_030229, GW190527\_092055, GW190708\_232457, GW190720\_000836, GW231028\_153006, and GW231123\_135430. Figure \ref{fig:sys} shows the systematics seen in these events when compared to the mixed and precessing (SEOBNRv4PHM or SEOBNRv5PHM) results from GWTC-2.1 and GWTC-4. To note, there is not a publicly available SEOBNRv4PHM result from GWTC-2.1 for GW190521\_030229. From these events, SEOBNRv5EHM tended to disagree the most with the quasi-circular waveforms in the effective spin parameter. The higher mass systems tended to reflect waveform systematics, two of which are among the highest mass events in this analysis.
Several of these events have been previously identified as events with waveform systematics \cite{2026PhRvD.113h3001X, ligo-o4a-cbc-catalog_intro, 2022ApJ...924...79E, LIGO-O3-GW190412}, some of which are further discussed in Malagon et al. \cite{2026arXiv260512818M}.

We find the most significant systematics with GW190521\_030229 and GW231123\_135430 when comparing the SEOBNRv5EHM result to a quasi-circular result, particularly in the effective spin and redshift parameters. For GW190521\_030229, the SEOBNRv5EHM is notably narrower in the effective spin and redshift compared to the mixed result which is more broadly constrained.
The SEOBNRv5EHM posteriors are also slightly shifted towards smaller values in the primary component mass.
In GW231123\_135430, we find strong disagreements across the secondary component mass, effective spin, and redshift parameters.
The SEOBNRv5EHM result recovers a significantly smaller secondary component mass compared to the quasi-circular waveforms, which are consistent with each other.
The SEOBNRv5EHM result is also shifted in the effective spin and redshift, with the distribution being narrower in the former and broader in the latter.
The mixed and precessing results are relatively consistent across these parameters, but the mixed result does exhibit a more diffuse distribution.

For GW190527\_092055, the SEOBNRv5EHM result is broader across the effective spin and redshift parameters compared to the mixed and precessing results, with a slight offset in the redshift distribution.
GW231028\_153006 has moderate systematics in the component masses where SEOBNRv5PHM recovers a bimodal shape in both, while the mixed result shows this shape in only the secondary component mass. By contrast, we do not find this structure in the SEOBNRv5EHM result.
GW190708\_232457, GW190720\_000836, and GW190412\_053044 are consistent except for modest differences in the primary component mass and effective spin parameters.

\onecolumngrid

\begingroup
\small
\setlength{\LTcapwidth}{\textwidth}
\setlength{\LTleft}{0pt}
\setlength{\LTright}{0pt}
\setlength{\tabcolsep}{6pt}
\renewcommand{\arraystretch}{1.1}
\begin{longtable}{lcccccc}
\caption{A table of GW BBH events where parameter inference performed with \textsc{SEOBNRv5EHM} waveform model, events are sorted by $\ln\mathrm{BF}_{10}$ against $\epsilon=0$, most eccentric first. The final column reports $\ln \mathrm{BF}_{10}=-\ln\!\left(0.5\,p(\epsilon=0\mid D)\right)$ using a reflection KDE estimate of $p(\epsilon=0\mid D)$. The values of 153 BBH ($\rm FAR < 1~yr^{-1}$) are reported as 68\% credible intervals.}\label{tab:event_summary}\\
\hline
Event & $m_{1,\mathrm{src}}$ & $m_{2,\mathrm{src}}$ & $\chi_{\mathrm{eff}}$ & $z$ & $\epsilon$ & $\ln \mathrm{BF}_{10}$ \\
\hline
\endfirsthead
\hline
Event & $m_{1,\mathrm{src}}$ & $m_{2,\mathrm{src}}$ & $\chi_{\mathrm{eff}}$ & $z$ & $\epsilon$ & $\ln \mathrm{BF}_{10}$ \\
\hline
\endhead
\hline
\multicolumn{7}{r}{\textit{Continued on next page}}\\
\endfoot
\hline
\endlastfoot
\texttt{\detokenize{GW200129_065458}} & \(33.17^{+2.31}_{-1.90}\) & \(28.20^{+1.88}_{-2.10}\) & \(0.077^{+0.055}_{-0.055}\) & \(0.165^{+0.036}_{-0.038}\) & \(0.119^{+0.0198}_{-0.0234}\) & \(33.084\) \\
\texttt{\detokenize{GW191204_171526}} & \(11.12^{+1.37}_{-0.88}\) & \(8.71^{+0.75}_{-0.93}\) & \(0.136^{+0.032}_{-0.025}\) & \(0.132^{+0.027}_{-0.034}\) & \(0.072^{+0.0154}_{-0.028}\) & \(-0.151\) \\
\texttt{\detokenize{GW190521_030229}} & \(85.30^{+11.24}_{-9.65}\) & \(60.99^{+8.57}_{-9.76}\) & \(0.064^{+0.149}_{-0.134}\) & \(0.781^{+0.135}_{-0.137}\) & \(0.206^{+0.163}_{-0.14}\) & \(-0.159\) \\
\texttt{\detokenize{GW231001_140220}} & \(70.58^{+16.74}_{-13.78}\) & \(40.97^{+10.50}_{-11.12}\) & \(0.067^{+0.218}_{-0.173}\) & \(0.785^{+0.296}_{-0.252}\) & \(0.197^{+0.12}_{-0.135}\) & \(-0.273\) \\
\texttt{\detokenize{GW200216_220804}} & \(47.45^{+9.53}_{-7.05}\) & \(32.82^{+6.80}_{-7.98}\) & \(0.062^{+0.212}_{-0.202}\) & \(0.758^{+0.226}_{-0.206}\) & \(0.199^{+0.168}_{-0.138}\) & \(-0.276\) \\
\texttt{\detokenize{GW230819_171910}} & \(65.37^{+16.52}_{-12.54}\) & \(35.86^{+10.84}_{-10.48}\) & \(-0.112^{+0.178}_{-0.228}\) & \(0.669^{+0.280}_{-0.218}\) & \(0.164^{+0.136}_{-0.112}\) & \(-0.377\) \\
\texttt{\detokenize{GW190805_211137}} & \(41.33^{+8.28}_{-6.58}\) & \(29.02^{+6.15}_{-5.57}\) & \(0.401^{+0.171}_{-0.235}\) & \(1.087^{+0.313}_{-0.297}\) & \(0.161^{+0.15}_{-0.111}\) & \(-0.444\) \\
\texttt{\detokenize{GW190413_134308}} & \(44.79^{+7.51}_{-5.98}\) & \(32.40^{+5.84}_{-6.13}\) & \(0.021^{+0.173}_{-0.169}\) & \(0.793^{+0.194}_{-0.194}\) & \(0.165^{+0.146}_{-0.113}\) & \(-0.444\) \\
\texttt{\detokenize{GW191127_050227}} & \(46.59^{+12.31}_{-8.88}\) & \(27.66^{+7.71}_{-7.34}\) & \(0.221^{+0.224}_{-0.221}\) & \(0.825^{+0.274}_{-0.234}\) & \(0.16^{+0.156}_{-0.11}\) & \(-0.481\) \\
\texttt{\detokenize{GW231221_135041}} & \(42.34^{+9.99}_{-7.19}\) & \(29.46^{+6.72}_{-6.43}\) & \(-0.018^{+0.203}_{-0.214}\) & \(0.795^{+0.291}_{-0.260}\) & \(0.154^{+0.167}_{-0.107}\) & \(-0.527\) \\
\texttt{\detokenize{GW240107_013215}} & \(53.56^{+13.74}_{-9.79}\) & \(33.93^{+9.87}_{-12.29}\) & \(0.284^{+0.203}_{-0.224}\) & \(0.932^{+0.333}_{-0.302}\) & \(0.157^{+0.162}_{-0.111}\) & \(-0.532\) \\
\texttt{\detokenize{GW190706_222641}} & \(61.49^{+9.48}_{-8.81}\) & \(39.63^{+7.28}_{-7.10}\) & \(0.303^{+0.153}_{-0.180}\) & \(0.817^{+0.190}_{-0.197}\) & \(0.128^{+0.118}_{-0.0873}\) & \(-0.663\) \\
\texttt{\detokenize{GW231123_135430}} & \(141.18^{+18.90}_{-16.88}\) & \(60.75^{+13.46}_{-9.66}\) & \(0.673^{+0.058}_{-0.037}\) & \(0.750^{+0.155}_{-0.180}\) & \(0.101^{+0.0838}_{-0.0657}\) & \(-0.673\) \\
\texttt{\detokenize{GW190701_203306}} & \(52.62^{+6.46}_{-5.21}\) & \(40.68^{+5.18}_{-6.44}\) & \(-0.047^{+0.145}_{-0.170}\) & \(0.395^{+0.082}_{-0.081}\) & \(0.129^{+0.123}_{-0.0887}\) & \(-0.683\) \\
\texttt{\detokenize{GW190727_060333}} & \(37.01^{+4.75}_{-3.84}\) & \(29.15^{+3.81}_{-4.09}\) & \(0.090^{+0.145}_{-0.142}\) & \(0.558^{+0.127}_{-0.126}\) & \(0.128^{+0.126}_{-0.0891}\) & \(-0.717\) \\
\texttt{\detokenize{GW230709_122727}} & \(43.66^{+8.64}_{-6.90}\) & \(30.27^{+7.13}_{-7.67}\) & \(0.064^{+0.186}_{-0.170}\) & \(0.730^{+0.252}_{-0.228}\) & \(0.135^{+0.158}_{-0.0941}\) & \(-0.731\) \\
\texttt{\detokenize{GW231223_032836}} & \(44.34^{+7.99}_{-6.33}\) & \(33.12^{+6.68}_{-6.71}\) & \(-0.170^{+0.186}_{-0.201}\) & \(0.707^{+0.238}_{-0.216}\) & \(0.127^{+0.125}_{-0.0885}\) & \(-0.749\) \\
\texttt{\detokenize{GW190719_215514}} & \(34.37^{+11.24}_{-7.12}\) & \(20.22^{+5.52}_{-4.70}\) & \(0.297^{+0.196}_{-0.195}\) & \(0.739^{+0.305}_{-0.228}\) & \(0.125^{+0.125}_{-0.0876}\) & \(-0.750\) \\
\texttt{\detokenize{GW191230_180458}} & \(45.67^{+6.84}_{-5.44}\) & \(35.25^{+5.61}_{-5.56}\) & \(0.000^{+0.161}_{-0.171}\) & \(0.816^{+0.190}_{-0.194}\) & \(0.122^{+0.117}_{-0.0842}\) & \(-0.756\) \\
\texttt{\detokenize{GW231004_232346}} & \(60.47^{+13.66}_{-11.26}\) & \(35.01^{+9.19}_{-8.37}\) & \(-0.071^{+0.173}_{-0.218}\) & \(0.734^{+0.280}_{-0.231}\) & \(0.116^{+0.137}_{-0.0799}\) & \(-0.771\) \\
\texttt{\detokenize{GW231005_021030}} & \(79.89^{+16.62}_{-14.12}\) & \(50.33^{+12.19}_{-10.84}\) & \(0.115^{+0.220}_{-0.198}\) & \(0.963^{+0.321}_{-0.288}\) & \(0.124^{+0.106}_{-0.0865}\) & \(-0.775\) \\
\texttt{\detokenize{GW230704_212616}} & \(88.20^{+21.62}_{-17.30}\) & \(51.22^{+17.38}_{-16.18}\) & \(0.273^{+0.212}_{-0.235}\) & \(0.983^{+0.383}_{-0.327}\) & \(0.112^{+0.0945}_{-0.0759}\) & \(-0.791\) \\
\texttt{\detokenize{GW230820_212515}} & \(58.95^{+13.81}_{-9.95}\) & \(36.07^{+10.25}_{-14.82}\) & \(0.159^{+0.191}_{-0.176}\) & \(0.649^{+0.232}_{-0.215}\) & \(0.119^{+0.161}_{-0.0835}\) & \(-0.806\) \\
\texttt{\detokenize{GW231230_170116}} & \(51.84^{+13.63}_{-9.59}\) & \(34.83^{+8.60}_{-7.99}\) & \(-0.189^{+0.195}_{-0.206}\) & \(0.895^{+0.335}_{-0.292}\) & \(0.117^{+0.139}_{-0.0814}\) & \(-0.813\) \\
\texttt{\detokenize{GW230630_125806}} & \(48.76^{+10.77}_{-9.11}\) & \(30.63^{+8.12}_{-7.08}\) & \(0.182^{+0.194}_{-0.196}\) & \(0.842^{+0.346}_{-0.281}\) & \(0.113^{+0.111}_{-0.079}\) & \(-0.820\) \\
\texttt{\detokenize{GW190731_140936}} & \(37.98^{+6.61}_{-5.24}\) & \(27.74^{+5.19}_{-5.19}\) & \(0.094^{+0.160}_{-0.157}\) & \(0.703^{+0.214}_{-0.206}\) & \(0.114^{+0.11}_{-0.0796}\) & \(-0.846\) \\
\texttt{\detokenize{GW230831_015414}} & \(41.34^{+9.79}_{-6.95}\) & \(30.78^{+7.56}_{-6.23}\) & \(0.039^{+0.173}_{-0.165}\) & \(0.810^{+0.307}_{-0.283}\) & \(0.112^{+0.115}_{-0.0781}\) & \(-0.863\) \\
\texttt{\detokenize{GW190915_235702}} & \(31.04^{+3.91}_{-3.05}\) & \(23.75^{+2.67}_{-2.83}\) & \(-0.038^{+0.107}_{-0.116}\) & \(0.351^{+0.070}_{-0.076}\) & \(0.0844^{+0.0465}_{-0.0528}\) & \(-0.868\) \\
\texttt{\detokenize{GW190413_052954}} & \(32.44^{+5.73}_{-4.33}\) & \(24.00^{+3.88}_{-3.84}\) & \(0.033^{+0.173}_{-0.173}\) & \(0.660^{+0.191}_{-0.170}\) & \(0.102^{+0.104}_{-0.0705}\) & \(-0.899\) \\
\texttt{\detokenize{GW190527_092055}} & \(37.85^{+10.63}_{-7.36}\) & \(25.05^{+7.92}_{-6.76}\) & \(0.119^{+0.192}_{-0.179}\) & \(0.629^{+0.304}_{-0.214}\) & \(0.119^{+0.154}_{-0.0864}\) & \(-0.908\) \\
\texttt{\detokenize{GW191109_010717}} & \(55.55^{+6.59}_{-6.05}\) & \(43.11^{+5.52}_{-5.43}\) & \(-0.231^{+0.173}_{-0.144}\) & \(0.402^{+0.134}_{-0.122}\) & \(0.103^{+0.09}_{-0.0708}\) & \(-0.913\) \\
\texttt{\detokenize{GW230922_040658}} & \(73.54^{+15.38}_{-11.52}\) & \(52.60^{+12.59}_{-13.41}\) & \(0.354^{+0.140}_{-0.158}\) & \(0.978^{+0.296}_{-0.303}\) & \(0.101^{+0.0846}_{-0.0697}\) & \(-0.921\) \\
\texttt{\detokenize{GW190602_175927}} & \(66.17^{+10.63}_{-8.90}\) & \(45.44^{+8.38}_{-9.54}\) & \(0.167^{+0.158}_{-0.157}\) & \(0.566^{+0.172}_{-0.143}\) & \(0.103^{+0.105}_{-0.0718}\) & \(-0.930\) \\
\texttt{\detokenize{GW231129_081745}} & \(44.28^{+9.36}_{-8.67}\) & \(22.96^{+5.89}_{-4.97}\) & \(0.034^{+0.159}_{-0.153}\) & \(0.635^{+0.266}_{-0.217}\) & \(0.102^{+0.112}_{-0.0706}\) & \(-0.936\) \\
\texttt{\detokenize{GW190929_012149}} & \(59.55^{+10.91}_{-11.02}\) & \(26.16^{+7.95}_{-6.39}\) & \(-0.008^{+0.139}_{-0.145}\) & \(0.669^{+0.256}_{-0.196}\) & \(0.0996^{+0.0985}_{-0.0693}\) & \(-0.945\) \\
\texttt{\detokenize{GW230806_204041}} & \(50.82^{+9.99}_{-8.03}\) & \(35.57^{+7.96}_{-7.64}\) & \(0.077^{+0.170}_{-0.164}\) & \(0.818^{+0.267}_{-0.248}\) & \(0.0995^{+0.105}_{-0.0687}\) & \(-0.946\) \\
\texttt{\detokenize{GW230707_124047}} & \(45.65^{+6.81}_{-5.48}\) & \(36.22^{+6.01}_{-5.54}\) & \(-0.029^{+0.151}_{-0.164}\) & \(0.698^{+0.185}_{-0.203}\) & \(0.0993^{+0.102}_{-0.0683}\) & \(-0.947\) \\
\texttt{\detokenize{GW231118_071402}} & \(41.45^{+8.30}_{-6.64}\) & \(28.55^{+6.33}_{-5.96}\) & \(0.136^{+0.184}_{-0.175}\) & \(0.714^{+0.261}_{-0.225}\) & \(0.106^{+0.116}_{-0.0743}\) & \(-0.955\) \\
\texttt{\detokenize{GW230712_090405}} & \(29.25^{+6.55}_{-6.09}\) & \(15.95^{+9.35}_{-6.51}\) & \(-0.046^{+0.164}_{-0.179}\) & \(0.458^{+0.239}_{-0.158}\) & \(0.0996^{+0.124}_{-0.0693}\) & \(-0.968\) \\
\texttt{\detokenize{GW231113_122623}} & \(38.94^{+8.64}_{-6.14}\) & \(26.57^{+5.24}_{-5.28}\) & \(0.344^{+0.157}_{-0.194}\) & \(0.579^{+0.193}_{-0.173}\) & \(0.0999^{+0.0994}_{-0.069}\) & \(-0.973\) \\
\texttt{\detokenize{GW230708_230935}} & \(64.08^{+11.33}_{-9.75}\) & \(40.25^{+8.45}_{-8.70}\) & \(0.021^{+0.173}_{-0.169}\) & \(0.574^{+0.174}_{-0.165}\) & \(0.102^{+0.114}_{-0.0715}\) & \(-0.976\) \\
\texttt{\detokenize{GW231119_075248}} & \(48.85^{+12.39}_{-9.13}\) & \(33.68^{+8.84}_{-7.84}\) & \(-0.001^{+0.164}_{-0.174}\) & \(0.991^{+0.368}_{-0.323}\) & \(0.1^{+0.106}_{-0.0696}\) & \(-0.982\) \\
\texttt{\detokenize{GW230814_061920}} & \(69.79^{+12.65}_{-12.05}\) & \(39.66^{+10.31}_{-9.55}\) & \(0.152^{+0.166}_{-0.171}\) & \(0.655^{+0.257}_{-0.197}\) & \(0.0959^{+0.098}_{-0.066}\) & \(-0.987\) \\
\texttt{\detokenize{GW230708_053705}} & \(28.86^{+4.72}_{-3.68}\) & \(22.28^{+3.47}_{-3.21}\) & \(0.073^{+0.151}_{-0.140}\) & \(0.555^{+0.171}_{-0.166}\) & \(0.0944^{+0.09}_{-0.0658}\) & \(-0.988\) \\
\texttt{\detokenize{GW191103_012549}} & \(11.76^{+3.25}_{-1.70}\) & \(7.66^{+1.25}_{-1.54}\) & \(0.183^{+0.082}_{-0.069}\) & \(0.202^{+0.059}_{-0.061}\) & \(0.0793^{+0.0583}_{-0.0508}\) & \(-0.989\) \\
\texttt{\detokenize{GW231127_165300}} & \(44.34^{+10.24}_{-7.86}\) & \(28.79^{+7.12}_{-6.75}\) & \(0.046^{+0.180}_{-0.175}\) & \(0.721^{+0.264}_{-0.243}\) & \(0.102^{+0.112}_{-0.0711}\) & \(-0.993\) \\
\texttt{\detokenize{GW190620_030421}} & \(53.96^{+9.51}_{-8.18}\) & \(34.05^{+6.90}_{-6.54}\) & \(0.397^{+0.121}_{-0.135}\) & \(0.576^{+0.155}_{-0.154}\) & \(0.092^{+0.092}_{-0.0635}\) & \(-1.016\) \\
\texttt{\detokenize{GW230803_033412}} & \(42.27^{+8.77}_{-7.04}\) & \(29.02^{+6.80}_{-6.13}\) & \(0.045^{+0.172}_{-0.176}\) & \(0.784^{+0.277}_{-0.241}\) & \(0.0913^{+0.0951}_{-0.063}\) & \(-1.034\) \\
\texttt{\detokenize{GW230601_224134}} & \(63.57^{+9.71}_{-8.64}\) & \(43.24^{+8.58}_{-8.59}\) & \(0.004^{+0.169}_{-0.171}\) & \(0.607^{+0.177}_{-0.180}\) & \(0.0889^{+0.0911}_{-0.0616}\) & \(-1.040\) \\
\texttt{\detokenize{GW230608_205047}} & \(48.16^{+7.97}_{-7.10}\) & \(30.70^{+6.96}_{-6.76}\) & \(0.046^{+0.147}_{-0.142}\) & \(0.547^{+0.173}_{-0.160}\) & \(0.0859^{+0.09}_{-0.0584}\) & \(-1.062\) \\
\texttt{\detokenize{GW231014_040532}} & \(20.23^{+3.39}_{-2.56}\) & \(14.99^{+2.28}_{-2.30}\) & \(0.180^{+0.130}_{-0.139}\) & \(0.428^{+0.131}_{-0.133}\) & \(0.0893^{+0.0887}_{-0.0625}\) & \(-1.064\) \\
\texttt{\detokenize{GW231223_202619}} & \(11.22^{+2.21}_{-1.22}\) & \(8.28^{+0.96}_{-1.24}\) & \(0.106^{+0.079}_{-0.070}\) & \(0.173^{+0.059}_{-0.055}\) & \(0.0825^{+0.0831}_{-0.0572}\) & \(-1.103\) \\
\texttt{\detokenize{GW190519_153544}} & \(59.54^{+7.85}_{-8.15}\) & \(38.10^{+6.86}_{-6.67}\) & \(0.382^{+0.107}_{-0.121}\) & \(0.602^{+0.162}_{-0.172}\) & \(0.0863^{+0.0844}_{-0.0597}\) & \(-1.112\) \\
\texttt{\detokenize{GW231018_233037}} & \(11.90^{+2.91}_{-1.99}\) & \(7.12^{+1.34}_{-1.32}\) & \(0.004^{+0.103}_{-0.077}\) & \(0.276^{+0.086}_{-0.087}\) & \(0.0829^{+0.072}_{-0.0572}\) & \(-1.113\) \\
\texttt{\detokenize{GW230729_082317}} & \(12.53^{+4.24}_{-2.17}\) & \(7.41^{+1.52}_{-1.71}\) & \(0.119^{+0.131}_{-0.095}\) & \(0.302^{+0.091}_{-0.092}\) & \(0.0839^{+0.0909}_{-0.0585}\) & \(-1.131\) \\
\texttt{\detokenize{GW190828_063405}} & \(31.72^{+3.24}_{-2.71}\) & \(26.13^{+2.72}_{-2.65}\) & \(0.197^{+0.092}_{-0.097}\) & \(0.387^{+0.076}_{-0.105}\) & \(0.0721^{+0.0435}_{-0.0465}\) & \(-1.137\) \\
\texttt{\detokenize{GW230928_215827}} & \(50.21^{+11.16}_{-9.44}\) & \(29.47^{+7.89}_{-6.84}\) & \(0.387^{+0.126}_{-0.156}\) & \(0.814^{+0.291}_{-0.253}\) & \(0.0829^{+0.0881}_{-0.0575}\) & \(-1.154\) \\
\texttt{\detokenize{GW230706_104333}} & \(16.13^{+2.82}_{-2.03}\) & \(11.57^{+1.61}_{-1.67}\) & \(0.176^{+0.082}_{-0.085}\) & \(0.342^{+0.100}_{-0.109}\) & \(0.0776^{+0.081}_{-0.0537}\) & \(-1.180\) \\
\texttt{\detokenize{GW230825_041334}} & \(40.95^{+9.29}_{-7.41}\) & \(26.48^{+6.85}_{-5.53}\) & \(0.297^{+0.150}_{-0.195}\) & \(0.805^{+0.305}_{-0.286}\) & \(0.0801^{+0.0803}_{-0.0555}\) & \(-1.190\) \\
\texttt{\detokenize{GW200209_085452}} & \(33.48^{+5.21}_{-4.11}\) & \(26.45^{+3.97}_{-3.90}\) & \(-0.064^{+0.150}_{-0.169}\) & \(0.674^{+0.179}_{-0.168}\) & \(0.0821^{+0.0896}_{-0.0569}\) & \(-1.198\) \\
\texttt{\detokenize{GW231213_111417}} & \(35.35^{+6.05}_{-4.64}\) & \(27.39^{+4.80}_{-4.40}\) & \(0.078^{+0.153}_{-0.141}\) & \(0.663^{+0.198}_{-0.197}\) & \(0.0788^{+0.0811}_{-0.0551}\) & \(-1.207\) \\
\texttt{\detokenize{GW230930_110730}} & \(34.25^{+7.10}_{-5.31}\) & \(24.57^{+4.94}_{-4.57}\) & \(0.038^{+0.154}_{-0.159}\) & \(0.754^{+0.243}_{-0.230}\) & \(0.0752^{+0.075}_{-0.0519}\) & \(-1.224\) \\
\texttt{\detokenize{GW231005_091549}} & \(28.36^{+5.11}_{-4.02}\) & \(21.27^{+3.82}_{-3.60}\) & \(-0.024^{+0.137}_{-0.149}\) & \(0.613^{+0.208}_{-0.186}\) & \(0.0752^{+0.0773}_{-0.0521}\) & \(-1.225\) \\
\texttt{\detokenize{GW231118_005626}} & \(18.83^{+3.23}_{-2.92}\) & \(11.40^{+2.01}_{-1.66}\) & \(0.360^{+0.058}_{-0.061}\) & \(0.390^{+0.101}_{-0.110}\) & \(0.0743^{+0.0705}_{-0.0515}\) & \(-1.231\) \\
\texttt{\detokenize{GW240104_164932}} & \(41.25^{+4.85}_{-4.13}\) & \(32.27^{+4.13}_{-4.50}\) & \(0.122^{+0.114}_{-0.115}\) & \(0.353^{+0.101}_{-0.101}\) & \(0.072^{+0.0704}_{-0.0498}\) & \(-1.238\) \\
\texttt{\detokenize{GW231223_075055}} & \(11.52^{+3.72}_{-1.92}\) & \(6.92^{+1.32}_{-1.56}\) & \(0.091^{+0.140}_{-0.091}\) & \(0.211^{+0.069}_{-0.064}\) & \(0.0739^{+0.0766}_{-0.0512}\) & \(-1.252\) \\
\texttt{\detokenize{GW230624_113103}} & \(25.81^{+5.87}_{-4.06}\) & \(16.44^{+2.87}_{-2.80}\) & \(0.149^{+0.142}_{-0.148}\) & \(0.361^{+0.123}_{-0.112}\) & \(0.0734^{+0.0752}_{-0.0505}\) & \(-1.254\) \\
\texttt{\detokenize{GW231206_233134}} & \(34.75^{+4.90}_{-3.85}\) & \(27.63^{+4.15}_{-3.93}\) & \(-0.087^{+0.132}_{-0.154}\) & \(0.545^{+0.150}_{-0.176}\) & \(0.0765^{+0.0763}_{-0.0533}\) & \(-1.261\) \\
\texttt{\detokenize{GW230606_004305}} & \(38.10^{+6.73}_{-5.43}\) & \(25.34^{+5.22}_{-5.06}\) & \(-0.129^{+0.167}_{-0.185}\) & \(0.456^{+0.135}_{-0.142}\) & \(0.0726^{+0.0744}_{-0.0505}\) & \(-1.262\) \\
\texttt{\detokenize{GW230920_071124}} & \(31.76^{+4.92}_{-4.04}\) & \(23.54^{+3.91}_{-3.87}\) & \(-0.011^{+0.129}_{-0.138}\) & \(0.495^{+0.161}_{-0.146}\) & \(0.0669^{+0.0668}_{-0.046}\) & \(-1.268\) \\
\texttt{\detokenize{GW230723_101834}} & \(16.49^{+3.20}_{-2.30}\) & \(10.78^{+1.80}_{-1.75}\) & \(-0.189^{+0.102}_{-0.091}\) & \(0.289^{+0.080}_{-0.086}\) & \(0.0708^{+0.07}_{-0.0488}\) & \(-1.278\) \\
\texttt{\detokenize{GW230630_234532}} & \(9.95^{+2.11}_{-1.29}\) & \(6.76^{+0.99}_{-1.11}\) & \(-0.058^{+0.089}_{-0.054}\) & \(0.199^{+0.064}_{-0.060}\) & \(0.0694^{+0.0589}_{-0.0476}\) & \(-1.281\) \\
\texttt{\detokenize{GW231102_071736}} & \(59.95^{+8.31}_{-7.64}\) & \(42.09^{+7.49}_{-7.21}\) & \(0.052^{+0.124}_{-0.120}\) & \(0.620^{+0.168}_{-0.171}\) & \(0.0725^{+0.0701}_{-0.0501}\) & \(-1.284\) \\
\texttt{\detokenize{GW190930_133541}} & \(12.25^{+4.40}_{-1.81}\) & \(7.77^{+1.30}_{-1.85}\) & \(0.126^{+0.133}_{-0.088}\) & \(0.156^{+0.041}_{-0.043}\) & \(0.0666^{+0.056}_{-0.0454}\) & \(-1.294\) \\
\texttt{\detokenize{GW190803_022701}} & \(36.31^{+5.77}_{-4.50}\) & \(27.17^{+4.21}_{-4.42}\) & \(0.016^{+0.155}_{-0.155}\) & \(0.597^{+0.161}_{-0.150}\) & \(0.071^{+0.0762}_{-0.0495}\) & \(-1.316\) \\
\texttt{\detokenize{GW230704_021211}} & \(32.86^{+7.20}_{-5.54}\) & \(19.72^{+4.20}_{-3.86}\) & \(-0.005^{+0.139}_{-0.154}\) & \(0.466^{+0.165}_{-0.150}\) & \(0.0684^{+0.068}_{-0.0474}\) & \(-1.319\) \\
\texttt{\detokenize{GW230805_034249}} & \(30.84^{+5.73}_{-4.55}\) & \(22.28^{+4.57}_{-4.35}\) & \(0.048^{+0.171}_{-0.169}\) & \(0.587^{+0.197}_{-0.178}\) & \(0.0708^{+0.0742}_{-0.0493}\) & \(-1.321\) \\
\texttt{\detokenize{GW190517_055101}} & \(32.75^{+5.30}_{-4.22}\) & \(23.69^{+3.71}_{-3.56}\) & \(0.570^{+0.100}_{-0.093}\) & \(0.493^{+0.148}_{-0.142}\) & \(0.0697^{+0.069}_{-0.0481}\) & \(-1.328\) \\
\texttt{\detokenize{GW230609_064958}} & \(35.48^{+5.96}_{-4.74}\) & \(25.50^{+4.62}_{-4.55}\) & \(-0.139^{+0.146}_{-0.156}\) & \(0.541^{+0.159}_{-0.162}\) & \(0.0728^{+0.0736}_{-0.0509}\) & \(-1.329\) \\
\texttt{\detokenize{GW230911_195324}} & \(32.07^{+4.79}_{-4.19}\) & \(22.78^{+3.02}_{-3.02}\) & \(0.018^{+0.122}_{-0.124}\) & \(0.283^{+0.115}_{-0.094}\) & \(0.0705^{+0.0727}_{-0.0492}\) & \(-1.340\) \\
\texttt{\detokenize{GW230824_033047}} & \(52.61^{+8.92}_{-7.56}\) & \(37.63^{+7.71}_{-7.41}\) & \(-0.005^{+0.143}_{-0.144}\) & \(0.719^{+0.227}_{-0.216}\) & \(0.069^{+0.0721}_{-0.048}\) & \(-1.345\) \\
\texttt{\detokenize{GW190513_205428}} & \(34.24^{+6.30}_{-6.09}\) & \(18.48^{+4.41}_{-3.27}\) & \(0.156^{+0.136}_{-0.141}\) & \(0.397^{+0.085}_{-0.087}\) & \(0.0627^{+0.0437}_{-0.0416}\) & \(-1.359\) \\
\texttt{\detokenize{GW200208_130117}} & \(36.17^{+5.25}_{-4.48}\) & \(27.12^{+4.10}_{-4.33}\) & \(-0.007^{+0.130}_{-0.141}\) & \(0.478^{+0.165}_{-0.115}\) & \(0.0624^{+0.0529}_{-0.0422}\) & \(-1.368\) \\
\texttt{\detokenize{GW151012_095443}} & \(24.62^{+9.89}_{-4.54}\) & \(14.26^{+2.98}_{-3.49}\) & \(0.216^{+0.179}_{-0.169}\) & \(0.222^{+0.075}_{-0.072}\) & \(0.0668^{+0.0677}_{-0.0463}\) & \(-1.379\) \\
\texttt{\detokenize{GW200219_094415}} & \(35.98^{+6.17}_{-4.63}\) & \(26.20^{+4.40}_{-4.66}\) & \(-0.059^{+0.147}_{-0.178}\) & \(0.627^{+0.158}_{-0.173}\) & \(0.0666^{+0.0797}_{-0.0467}\) & \(-1.388\) \\
\texttt{\detokenize{GW240109_050431}} & \(28.57^{+4.59}_{-4.09}\) & \(18.29^{+3.03}_{-2.77}\) & \(-0.071^{+0.126}_{-0.137}\) & \(0.276^{+0.104}_{-0.089}\) & \(0.0646^{+0.0683}_{-0.0449}\) & \(-1.404\) \\
\texttt{\detokenize{GW230605_065343}} & \(17.69^{+4.38}_{-2.91}\) & \(10.72^{+1.88}_{-1.90}\) & \(0.068^{+0.108}_{-0.073}\) & \(0.209^{+0.070}_{-0.061}\) & \(0.064^{+0.0622}_{-0.0445}\) & \(-1.413\) \\
\texttt{\detokenize{GW231008_142521}} & \(43.52^{+9.27}_{-7.57}\) & \(25.40^{+6.60}_{-6.29}\) & \(-0.010^{+0.152}_{-0.166}\) & \(0.554^{+0.207}_{-0.163}\) & \(0.0634^{+0.0675}_{-0.044}\) & \(-1.421\) \\
\texttt{\detokenize{GW230726_002940}} & \(35.60^{+4.96}_{-3.94}\) & \(27.94^{+3.61}_{-3.52}\) & \(-0.013^{+0.125}_{-0.139}\) & \(0.364^{+0.117}_{-0.110}\) & \(0.0622^{+0.0627}_{-0.0435}\) & \(-1.428\) \\
\texttt{\detokenize{GW191215_223052}} & \(23.91^{+3.54}_{-2.64}\) & \(17.73^{+2.17}_{-2.35}\) & \(-0.004^{+0.099}_{-0.098}\) & \(0.407^{+0.080}_{-0.097}\) & \(0.0581^{+0.0502}_{-0.0395}\) & \(-1.461\) \\
\texttt{\detokenize{GW231028_153006}} & \(82.19^{+10.69}_{-7.03}\) & \(67.96^{+7.84}_{-8.11}\) & \(0.489^{+0.092}_{-0.079}\) & \(0.708^{+0.098}_{-0.147}\) & \(0.0602^{+0.0499}_{-0.0414}\) & \(-1.464\) \\
\texttt{\detokenize{GW231113_200417}} & \(11.02^{+2.42}_{-1.41}\) & \(7.59^{+1.10}_{-1.31}\) & \(0.120^{+0.071}_{-0.049}\) & \(0.235^{+0.075}_{-0.071}\) & \(0.0616^{+0.0581}_{-0.0425}\) & \(-1.482\) \\
\texttt{\detokenize{GW230702_185453}} & \(42.18^{+14.34}_{-11.03}\) & \(16.62^{+5.84}_{-4.08}\) & \(0.082^{+0.213}_{-0.178}\) & \(0.467^{+0.165}_{-0.140}\) & \(0.0611^{+0.0636}_{-0.0423}\) & \(-1.483\) \\
\texttt{\detokenize{GW231114_043211}} & \(24.89^{+6.39}_{-5.03}\) & \(7.45^{+1.61}_{-1.31}\) & \(0.138^{+0.144}_{-0.132}\) & \(0.275^{+0.085}_{-0.077}\) & \(0.0591^{+0.0618}_{-0.0412}\) & \(-1.497\) \\
\texttt{\detokenize{GW230904_051013}} & \(10.61^{+2.43}_{-1.47}\) & \(6.99^{+1.09}_{-1.23}\) & \(0.045^{+0.088}_{-0.053}\) & \(0.208^{+0.070}_{-0.062}\) & \(0.0571^{+0.0511}_{-0.0396}\) & \(-1.500\) \\
\texttt{\detokenize{GW231029_111508}} & \(63.30^{+10.21}_{-9.26}\) & \(41.89^{+8.81}_{-9.18}\) & \(0.139^{+0.143}_{-0.141}\) & \(0.522^{+0.191}_{-0.168}\) & \(0.0585^{+0.0606}_{-0.0407}\) & \(-1.508\) \\
\texttt{\detokenize{GW231224_024321}} & \(9.32^{+1.17}_{-0.79}\) & \(7.33^{+0.66}_{-0.81}\) & \(-0.003^{+0.038}_{-0.032}\) & \(0.185^{+0.039}_{-0.052}\) & \(0.0586^{+0.0554}_{-0.0406}\) & \(-1.513\) \\
\texttt{\detokenize{GW190421_213856}} & \(41.14^{+6.26}_{-4.91}\) & \(30.69^{+4.80}_{-5.05}\) & \(-0.089^{+0.149}_{-0.163}\) & \(0.496^{+0.140}_{-0.143}\) & \(0.0588^{+0.0619}_{-0.0411}\) & \(-1.520\) \\
\texttt{\detokenize{GW191222_033537}} & \(43.36^{+6.27}_{-5.10}\) & \(33.10^{+5.32}_{-5.33}\) & \(-0.057^{+0.125}_{-0.143}\) & \(0.548^{+0.143}_{-0.167}\) & \(0.0563^{+0.0527}_{-0.0393}\) & \(-1.539\) \\
\texttt{\detokenize{GW190521_074359}} & \(42.13^{+3.44}_{-3.38}\) & \(32.87^{+3.24}_{-3.39}\) & \(0.060^{+0.074}_{-0.072}\) & \(0.227^{+0.054}_{-0.058}\) & \(0.045^{+0.028}_{-0.029}\) & \(-1.552\) \\
\texttt{\detokenize{GW191105_143521}} & \(10.76^{+2.21}_{-1.30}\) & \(7.54^{+1.01}_{-1.23}\) & \(-0.026^{+0.079}_{-0.057}\) & \(0.226^{+0.051}_{-0.059}\) & \(0.0523^{+0.0513}_{-0.0364}\) & \(-1.557\) \\
\texttt{\detokenize{GW230927_043729}} & \(34.56^{+5.04}_{-4.11}\) & \(26.79^{+4.25}_{-3.92}\) & \(0.010^{+0.127}_{-0.130}\) & \(0.551^{+0.155}_{-0.168}\) & \(0.0557^{+0.0561}_{-0.0388}\) & \(-1.561\) \\
\texttt{\detokenize{GW190828_065509}} & \(23.66^{+4.73}_{-4.57}\) & \(10.11^{+2.19}_{-1.50}\) & \(0.076^{+0.109}_{-0.103}\) & \(0.315^{+0.078}_{-0.085}\) & \(0.054^{+0.0541}_{-0.0373}\) & \(-1.582\) \\
\texttt{\detokenize{GW200302_015811}} & \(36.61^{+5.69}_{-5.70}\) & \(19.41^{+4.93}_{-3.67}\) & \(0.020^{+0.143}_{-0.143}\) & \(0.306^{+0.099}_{-0.091}\) & \(0.0528^{+0.0531}_{-0.0369}\) & \(-1.583\) \\
\texttt{\detokenize{GW231108_125142}} & \(23.16^{+3.15}_{-2.48}\) & \(17.54^{+2.03}_{-2.05}\) & \(-0.065^{+0.080}_{-0.087}\) & \(0.366^{+0.082}_{-0.097}\) & \(0.0526^{+0.0537}_{-0.0367}\) & \(-1.584\) \\
\texttt{\detokenize{GW190925_232845}} & \(20.84^{+3.55}_{-2.17}\) & \(15.43^{+1.84}_{-2.23}\) & \(0.091^{+0.094}_{-0.090}\) & \(0.186^{+0.053}_{-0.049}\) & \(0.0531^{+0.0507}_{-0.0367}\) & \(-1.586\) \\
\texttt{\detokenize{GW230922_020344}} & \(38.58^{+5.09}_{-4.04}\) & \(29.29^{+3.37}_{-3.58}\) & \(0.057^{+0.123}_{-0.116}\) & \(0.317^{+0.069}_{-0.079}\) & \(0.0537^{+0.0531}_{-0.0374}\) & \(-1.592\) \\
\texttt{\detokenize{GW190910_112807}} & \(41.70^{+4.95}_{-4.17}\) & \(32.79^{+3.80}_{-4.10}\) & \(-0.016^{+0.118}_{-0.124}\) & \(0.354^{+0.095}_{-0.108}\) & \(0.0526^{+0.0505}_{-0.0364}\) & \(-1.595\) \\
\texttt{\detokenize{GW190708_232457}} & \(17.31^{+2.71}_{-1.62}\) & \(13.13^{+1.36}_{-1.80}\) & \(0.009^{+0.060}_{-0.049}\) & \(0.182^{+0.041}_{-0.048}\) & \(0.0492^{+0.0374}_{-0.0331}\) & \(-1.597\) \\
\texttt{\detokenize{GW200128_022011}} & \(38.48^{+5.85}_{-4.67}\) & \(30.00^{+4.87}_{-4.36}\) & \(0.120^{+0.130}_{-0.137}\) & \(0.688^{+0.176}_{-0.191}\) & \(0.0528^{+0.0542}_{-0.0367}\) & \(-1.603\) \\
\texttt{\detokenize{GW231110_040320}} & \(18.89^{+3.65}_{-2.56}\) & \(12.87^{+1.95}_{-2.02}\) & \(0.179^{+0.080}_{-0.075}\) & \(0.347^{+0.096}_{-0.107}\) & \(0.0518^{+0.0529}_{-0.0358}\) & \(-1.626\) \\
\texttt{\detokenize{GW230924_124453}} & \(28.50^{+3.24}_{-2.69}\) & \(23.23^{+2.75}_{-2.64}\) & \(0.019^{+0.103}_{-0.109}\) & \(0.413^{+0.106}_{-0.113}\) & \(0.0489^{+0.0497}_{-0.0338}\) & \(-1.633\) \\
\texttt{\detokenize{GW190725_174728}} & \(11.88^{+3.64}_{-2.46}\) & \(6.06^{+1.42}_{-1.25}\) & \(-0.031^{+0.169}_{-0.099}\) & \(0.225^{+0.064}_{-0.065}\) & \(0.0474^{+0.0452}_{-0.0321}\) & \(-1.646\) \\
\texttt{\detokenize{GW231231_154016}} & \(22.48^{+3.14}_{-2.21}\) & \(17.32^{+1.83}_{-2.02}\) & \(-0.044^{+0.069}_{-0.072}\) & \(0.197^{+0.072}_{-0.063}\) & \(0.0493^{+0.0509}_{-0.0342}\) & \(-1.669\) \\
\texttt{\detokenize{GW200224_222234}} & \(39.10^{+3.62}_{-2.99}\) & \(32.04^{+2.85}_{-3.49}\) & \(0.081^{+0.093}_{-0.095}\) & \(0.315^{+0.051}_{-0.066}\) & \(0.0482^{+0.0398}_{-0.0331}\) & \(-1.673\) \\
\texttt{\detokenize{GW230731_215307}} & \(10.47^{+1.65}_{-1.05}\) & \(7.81^{+0.88}_{-1.05}\) & \(-0.041^{+0.056}_{-0.040}\) & \(0.208^{+0.048}_{-0.057}\) & \(0.0493^{+0.049}_{-0.0344}\) & \(-1.676\) \\
\texttt{\detokenize{GW231104_133418}} & \(12.43^{+2.69}_{-1.59}\) & \(8.59^{+1.24}_{-1.47}\) & \(0.142^{+0.065}_{-0.047}\) & \(0.263^{+0.069}_{-0.076}\) & \(0.0491^{+0.0513}_{-0.0339}\) & \(-1.681\) \\
\texttt{\detokenize{GW230628_231200}} & \(32.25^{+3.21}_{-2.62}\) & \(27.26^{+2.83}_{-2.83}\) & \(-0.032^{+0.094}_{-0.108}\) & \(0.384^{+0.083}_{-0.108}\) & \(0.0478^{+0.047}_{-0.033}\) & \(-1.692\) \\
\texttt{\detokenize{GW170729_185629}} & \(50.11^{+7.64}_{-7.70}\) & \(29.37^{+6.66}_{-5.59}\) & \(0.272^{+0.146}_{-0.186}\) & \(0.474^{+0.146}_{-0.139}\) & \(0.0504^{+0.0504}_{-0.0356}\) & \(-1.704\) \\
\texttt{\detokenize{GW200225_060421}} & \(18.42^{+2.24}_{-1.69}\) & \(14.43^{+1.50}_{-1.71}\) & \(-0.087^{+0.105}_{-0.153}\) & \(0.252^{+0.056}_{-0.068}\) & \(0.0466^{+0.0472}_{-0.0324}\) & \(-1.720\) \\
\texttt{\detokenize{GW170823_131358}} & \(38.06^{+5.40}_{-4.13}\) & \(28.69^{+3.84}_{-4.17}\) & \(0.087^{+0.135}_{-0.132}\) & \(0.366^{+0.086}_{-0.095}\) & \(0.0457^{+0.0482}_{-0.0318}\) & \(-1.742\) \\
\texttt{\detokenize{GW230811_032116}} & \(33.82^{+4.91}_{-4.51}\) & \(23.21^{+3.67}_{-3.23}\) & \(0.030^{+0.109}_{-0.106}\) & \(0.388^{+0.124}_{-0.129}\) & \(0.0459^{+0.0465}_{-0.0317}\) & \(-1.747\) \\
\texttt{\detokenize{GW231020_142947}} & \(12.63^{+5.91}_{-2.41}\) & \(7.10^{+1.61}_{-2.07}\) & \(0.152^{+0.177}_{-0.093}\) & \(0.226^{+0.065}_{-0.077}\) & \(0.0484^{+0.0479}_{-0.0339}\) & \(-1.750\) \\
\texttt{\detokenize{GW230914_111401}} & \(59.35^{+7.14}_{-7.48}\) & \(35.72^{+8.32}_{-7.98}\) & \(0.135^{+0.118}_{-0.118}\) & \(0.475^{+0.136}_{-0.124}\) & \(0.0432^{+0.0419}_{-0.0299}\) & \(-1.752\) \\
\texttt{\detokenize{GW200202_154313}} & \(10.55^{+1.71}_{-1.06}\) & \(7.03^{+0.73}_{-0.90}\) & \(0.034^{+0.070}_{-0.048}\) & \(0.088^{+0.021}_{-0.023}\) & \(0.0419^{+0.0371}_{-0.028}\) & \(-1.763\) \\
\texttt{\detokenize{GW170818_022509}} & \(33.48^{+3.72}_{-2.98}\) & \(26.72^{+2.79}_{-3.17}\) & \(-0.023^{+0.111}_{-0.124}\) & \(0.251^{+0.075}_{-0.056}\) & \(0.0446^{+0.0451}_{-0.0308}\) & \(-1.771\) \\
\texttt{\detokenize{GW231118_090602}} & \(12.34^{+4.27}_{-1.98}\) & \(7.64^{+1.34}_{-1.75}\) & \(0.070^{+0.134}_{-0.056}\) & \(0.255^{+0.068}_{-0.073}\) & \(0.0434^{+0.0446}_{-0.03}\) & \(-1.783\) \\
\texttt{\detokenize{GW230919_215712}} & \(26.79^{+3.06}_{-2.45}\) & \(21.06^{+2.34}_{-2.53}\) & \(0.172^{+0.075}_{-0.077}\) & \(0.270^{+0.089}_{-0.070}\) & \(0.0432^{+0.0441}_{-0.03}\) & \(-1.791\) \\
\texttt{\detokenize{GW190503_185404}} & \(40.81^{+6.12}_{-5.17}\) & \(28.10^{+4.54}_{-5.02}\) & \(-0.018^{+0.135}_{-0.168}\) & \(0.301^{+0.073}_{-0.076}\) & \(0.0442^{+0.045}_{-0.0309}\) & \(-1.793\) \\
\texttt{\detokenize{GW190408_181802}} & \(24.78^{+3.16}_{-2.53}\) & \(18.38^{+2.17}_{-2.35}\) & \(-0.043^{+0.089}_{-0.096}\) & \(0.287^{+0.053}_{-0.071}\) & \(0.0444^{+0.043}_{-0.031}\) & \(-1.798\) \\
\texttt{\detokenize{GW190720_000836}} & \(12.83^{+3.42}_{-2.04}\) & \(7.89^{+1.43}_{-1.51}\) & \(0.148^{+0.092}_{-0.061}\) & \(0.173^{+0.086}_{-0.044}\) & \(0.0397^{+0.0373}_{-0.0272}\) & \(-1.844\) \\
\texttt{\detokenize{GW200316_215756}} & \(14.56^{+9.79}_{-3.47}\) & \(6.94^{+1.84}_{-2.29}\) & \(0.158^{+0.252}_{-0.104}\) & \(0.234^{+0.063}_{-0.068}\) & \(0.0416^{+0.0421}_{-0.0288}\) & \(-1.847\) \\
\texttt{\detokenize{GW190924_021846}} & \(9.10^{+2.92}_{-1.73}\) & \(4.92^{+1.04}_{-1.02}\) & \(0.023^{+0.155}_{-0.074}\) & \(0.112^{+0.029}_{-0.031}\) & \(0.0391^{+0.0318}_{-0.0269}\) & \(-1.858\) \\
\texttt{\detokenize{GW190707_093326}} & \(11.22^{+1.69}_{-1.08}\) & \(8.47^{+0.89}_{-1.08}\) & \(-0.064^{+0.057}_{-0.045}\) & \(0.170^{+0.042}_{-0.050}\) & \(0.0393^{+0.0429}_{-0.0276}\) & \(-1.896\) \\
\texttt{\detokenize{GW191129_134029}} & \(10.77^{+2.67}_{-1.63}\) & \(6.72^{+1.12}_{-1.20}\) & \(0.065^{+0.102}_{-0.053}\) & \(0.155^{+0.037}_{-0.044}\) & \(0.0379^{+0.0353}_{-0.026}\) & \(-1.906\) \\
\texttt{\detokenize{GW190512_180714}} & \(22.44^{+4.06}_{-3.76}\) & \(12.70^{+2.32}_{-1.85}\) & \(0.028^{+0.087}_{-0.082}\) & \(0.289^{+0.061}_{-0.076}\) & \(0.0389^{+0.0389}_{-0.027}\) & \(-1.917\) \\
\texttt{\detokenize{GW231206_233901}} & \(37.61^{+3.81}_{-3.34}\) & \(27.95^{+3.67}_{-3.76}\) & \(-0.040^{+0.085}_{-0.088}\) & \(0.285^{+0.037}_{-0.050}\) & \(0.0368^{+0.036}_{-0.0254}\) & \(-1.960\) \\
\texttt{\detokenize{GW170809_082821}} & \(33.22^{+4.52}_{-3.39}\) & \(24.56^{+2.95}_{-3.45}\) & \(0.071^{+0.121}_{-0.111}\) & \(0.213^{+0.041}_{-0.047}\) & \(0.0363^{+0.0366}_{-0.0253}\) & \(-2.000\) \\
\texttt{\detokenize{GW170814_103043}} & \(31.06^{+3.49}_{-2.56}\) & \(24.45^{+2.12}_{-2.45}\) & \(0.089^{+0.074}_{-0.076}\) & \(0.128^{+0.026}_{-0.031}\) & \(0.035^{+0.0355}_{-0.0244}\) & \(-2.014\) \\
\texttt{\detokenize{GW190728_064510}} & \(12.14^{+4.09}_{-1.60}\) & \(8.18^{+1.19}_{-1.84}\) & \(0.119^{+0.115}_{-0.049}\) & \(0.175^{+0.036}_{-0.049}\) & \(0.0343^{+0.0333}_{-0.0238}\) & \(-2.040\) \\
\texttt{\detokenize{GW170104_101158}} & \(28.40^{+3.61}_{-2.76}\) & \(21.19^{+2.55}_{-2.92}\) & \(-0.014^{+0.099}_{-0.115}\) & \(0.220^{+0.050}_{-0.061}\) & \(0.0352^{+0.0364}_{-0.0246}\) & \(-2.044\) \\
\texttt{\detokenize{GW151226_033853}} & \(12.69^{+3.23}_{-1.82}\) & \(8.16^{+1.31}_{-1.51}\) & \(0.183^{+0.078}_{-0.048}\) & \(0.100^{+0.026}_{-0.030}\) & \(0.0343^{+0.0321}_{-0.0239}\) & \(-2.075\) \\
\texttt{\detokenize{GW230927_153832}} & \(21.53^{+2.26}_{-1.86}\) & \(16.95^{+1.60}_{-1.76}\) & \(0.041^{+0.045}_{-0.043}\) & \(0.225^{+0.052}_{-0.067}\) & \(0.0318^{+0.0338}_{-0.0222}\) & \(-2.088\) \\
\texttt{\detokenize{GW200311_115853}} & \(33.52^{+3.31}_{-2.46}\) & \(27.43^{+2.41}_{-3.13}\) & \(-0.045^{+0.098}_{-0.118}\) & \(0.229^{+0.034}_{-0.045}\) & \(0.0294^{+0.0304}_{-0.0205}\) & \(-2.187\) \\
\texttt{\detokenize{GW190630_185205}} & \(34.10^{+4.94}_{-4.19}\) & \(22.53^{+3.77}_{-3.45}\) & \(0.082^{+0.093}_{-0.094}\) & \(0.215^{+0.058}_{-0.063}\) & \(0.0288^{+0.0289}_{-0.0199}\) & \(-2.196\) \\
\texttt{\detokenize{GW190412_053044}} & \(30.29^{+2.29}_{-2.12}\) & \(8.17^{+0.55}_{-0.50}\) & \(0.236^{+0.049}_{-0.050}\) & \(0.152^{+0.025}_{-0.027}\) & \(0.0285^{+0.0279}_{-0.0199}\) & \(-2.211\) \\
\texttt{\detokenize{GW200112_155838}} & \(35.96^{+3.92}_{-3.21}\) & \(27.79^{+3.01}_{-3.42}\) & \(0.052^{+0.098}_{-0.095}\) & \(0.234^{+0.052}_{-0.055}\) & \(0.0275^{+0.0291}_{-0.019}\) & \(-2.237\) \\
\texttt{\detokenize{GW191216_213338}} & \(12.70^{+3.03}_{-2.07}\) & \(7.33^{+1.28}_{-1.24}\) & \(0.125^{+0.091}_{-0.051}\) & \(0.073^{+0.017}_{-0.016}\) & \(0.0279^{+0.0263}_{-0.0196}\) & \(-2.289\) \\
\texttt{\detokenize{GW170608_020116}} & \(10.57^{+1.86}_{-1.04}\) & \(7.85^{+0.84}_{-1.09}\) & \(0.041^{+0.058}_{-0.033}\) & \(0.073^{+0.019}_{-0.019}\) & \(0.0261^{+0.0273}_{-0.0181}\) & \(-2.297\) \\
\texttt{\detokenize{GW150914_095045}} & \(34.34^{+2.28}_{-1.74}\) & \(30.04^{+1.72}_{-2.15}\) & \(-0.058^{+0.074}_{-0.079}\) & \(0.101^{+0.020}_{-0.022}\) & \(0.0242^{+0.0246}_{-0.0167}\) & \(-2.343\) \\
\texttt{\detokenize{GW231226_101520}} & \(39.36^{+2.37}_{-1.85}\) & \(34.86^{+1.97}_{-2.66}\) & \(-0.108^{+0.061}_{-0.055}\) & \(0.231^{+0.028}_{-0.041}\) & \(0.0233^{+0.0233}_{-0.0163}\) & \(-2.468\) \\
\texttt{\detokenize{GW230627_015337}} & \(8.39^{+1.07}_{-0.95}\) & \(5.81^{+0.70}_{-0.63}\) & \(0.016^{+0.047}_{-0.025}\) & \(0.062^{+0.012}_{-0.018}\) & \(0.02^{+0.0193}_{-0.0138}\) & \(-2.551\) \\
\texttt{\detokenize{GW230814_230901}} & \(34.30^{+2.22}_{-1.89}\) & \(27.69^{+1.83}_{-2.20}\) & \(0.014^{+0.038}_{-0.035}\) & \(0.071^{+0.026}_{-0.021}\) & \(0.0185^{+0.0188}_{-0.0129}\) & \(-2.679\) \\
\end{longtable}
\endgroup

\twocolumngrid

\bibstyle{unsrt}
\bibliography{reference}

\end{document}